\documentclass[aps,preprint,longbibliography,graphicx]{revtex4-1}

\usepackage{graphicx}
\usepackage{epstopdf}
\usepackage{caption}
\usepackage{subcaption}
\usepackage{mathtools}
\usepackage[percent]{overpic}
\usepackage[export]{adjustbox}
\usepackage[colorlinks=true,linkcolor=blue]{hyperref}%
\usepackage{soul}

\draft

\usepackage{color}

\begin{document}

\title{Luminescence from cavitation bubbles deformed in uniform pressure gradients} 
\date{today}

\author{Outi Supponen}
\affiliation{Laboratory for Hydraulic Machines, Ecole Polytechnique F\'ed\'erale de Lausanne, Avenue de Cour 33 Bis, 1007 Lausanne, Switzerland}
\author{Danail Obreschkow}
\affiliation{International Centre for Radio Astronomy Research, University of Western Australia, M468 7 Fairway, Crawley, Western Australia 6009, Australia}
\author{Philippe Kobel}
\affiliation{Laboratory for Hydraulic Machines, Ecole Polytechnique F\'ed\'erale de Lausanne, Avenue de Cour 33 Bis, 1007 Lausanne, Switzerland}
\author{Mohamed Farhat}
\affiliation{Laboratory for Hydraulic Machines, Ecole Polytechnique F\'ed\'erale de Lausanne, Avenue de Cour 33 Bis, 1007 Lausanne, Switzerland}

\date{\today}


\begin{abstract}
Presented here are observations that demonstrate how the deformation of millimetric cavitation bubbles by a uniform pressure gradient quenches single collapse luminescence.
Our innovative measurement system captures a broad luminescence spectrum (wavelength range $300$--$900$~nm) from the individual collapses of laser-induced bubbles in water.
By varying the bubble size, driving pressure and the perceived gravity level aboard parabolic flights, we probed the limit from aspherical to highly spherical bubble collapses. 
Luminescence was detected for bubbles of maximum radii within the previously uncovered range $R_{0} =1.5$--6~mm for laser-induced bubbles.
The relative luminescence energy was found to rapidly decrease as a function of bubble asymmetry quantified by the anisotropy parameter $\zeta$, which is the dimensionless equivalent of the Kelvin impulse. 
As established previously, $\zeta$ also dictates the characteristic parameters of bubble-driven microjets.
The threshold of $\zeta$ beyond which no luminescence is observed in our experiment closely coincides with the threshold where the microjets visibly pierce the bubble and drive a vapor-jet during the rebound.
The individual fitted blackbody temperatures range between $T_{\rm lum}=7000$ and 11500~K but do not show any clear trend as a function of $\zeta$.
Time-resolved measurements using a high-speed photodetector disclose multiple luminescence events at each bubble collapse.
The averaged full width at half maximum of the pulse is found to scale with $R_{0}$ and to range between 10--20~ns.
\end{abstract}

\maketitle

\section{Introduction}
As a cavitation bubble undergoes a spherical collapse, it compresses its enclosed gaseous contents and - presumably - adiabatically heats them to temperatures of several thousands of degrees, which results in light emission called luminescence~\cite{Brenner2002}. 
The drive to investigate luminescence comes from the intense energy focusing at a bubble collapse that provides a catalytic host for unique chemical reactions~\cite{Didenko1999,Didenko2002}, offering a potential for cancer therapy~\cite{Rosenthal2004,Yu2004}, environmental remediation~\cite{Mason2007,Adewuyi2005} and fabrication of nanomaterials~\cite{Suslick1999,Gedanken2004}.
While most past studies have researched sonoluminescence, that is, luminescence from acoustically driven bubbles, light emission has also been detected from hydrodynamic cavitation in engineering flows~\cite{Leighton2003,Farhat2010}.

Due to the occurrence at the last instant of the collapse, the redistribution of the bubble's energy into luminescence, as well as shock waves, microjets and elastic rebounds~(see introduction in Ref.~\cite{Obreschkow2013}), must be highly sensitive to the topological changes of the cavity volume during the final collapse stage. 
This represents an important feature, considering that any anisotropy in the pressure field of the surrounding liquid will result in a deformation of an initially spherical bubble, inducing a microjet that pierces the bubble and therefore making it undergo a toroidal collapse~\cite{Supponen2015,Supponen2016}. 
The level of compression of the bubble gases is reduced for even slight bubble deformations, manifested in the weakening of the collapse shock wave emissions~\cite{Vogel1988,Supponen2017}.
Indeed, luminescence has been shown to vary with the proximity of near surfaces that break the spherical symmetry of the bubble~\cite{Ohl1998,Ohl2002,Brujan2005b}. 
It has also been shown that the lack of buoyancy enhances the energy concentration at the final stage of the bubble collapse~\cite{Matula2000}, even for bubbles that are highly spherical and generally assumed not to be subject to deformation by gravity (maximum bubble radius $R_{0} \sim 40$~$\mu$m at atmospheric pressure). 
Bubbles collapsing with pronounced microjets in multibubble fields have been shown to emit less light (or none) compared to the spherically collapsing bubbles~\cite{Cairos2017}. 

Spectral analyses on luminescence have proposed a wide range of temperatures at the bubble collapse in water, depending on whether the bubble is trapped in an oscillating acoustic field (bubble temperatures $T > 10^{4}$~K)~\cite{Brenner2002}, induced by a laser pulse ($T \sim 7000$--$8000$~K)~\cite{Brujan2005,Baghda2001}, induced by a spark ($T\sim6700$~K)~\cite{Zhang2017} or within a bubble cloud ($T < 5000$~K)~\cite{McNamara1999,Didenko1999}. 
Recent studies reached $1.4\times10^{4}$~K for an energetic bubble collapse provoked by piezo-electric tranducers~\cite{Ramsey2013} and over $2\times10^{4}$~K for a centimetric bubble expanded by a chemical reaction in a free fall tower~\cite{Duplat2015}.
Moreover, luminescence spectra from small bubbles (maximum radius $R_{0}<1$~mm) show a smooth continuum similar to a blackbody, while spectra of luminescence from large, laser-induced bubbles ($R_{0}>1$~mm) and multibubble sonoluminescence (MBSL) have shown emission lines of excited hydroxyl (OH$^{-}$) bands at 310~nm~\cite{Matula1995,Brujan2005b} that has been associated with aspherical bubble collapses.
It is unclear, however, to what extent the spectral differences in these distinct scenarios are caused by physical or experimental reasons, and a systematic picture of the role of pressure field anisotropies - and the resulting bubble deformation - on luminescence is still lacking.

This work presents observations on the luminescence of initially highly spherical, millimetric bubbles collapsing at different levels of deformation caused by the gravity-induced uniform pressure gradient.
We probe the transition from toroidal jetting bubbles in controlled pressure gradients to highly spherical bubbles in microgravity and cover a broad parameter space.
Spectral and time-resolved measurements are made on single cavitation bubble luminescence (SCBL) from individual collapses of transient, laser-induced vapor bubbles in water, contrasting with the established single bubble sonoluminecence (SBSL), which is normally understood as the time-averaged light emitted by an oscillating bubble trapped in an acoustic field.
It also differs from the averaged SCBL, from luminescence of gas bubbles and from bubbles in liquids doped with noble gas.

\section{Experimental setup}

\begin{figure}
\begin{center}
\includegraphics[width=0.6\textwidth, trim=0cm 0cm 0.1cm 0cm, clip]{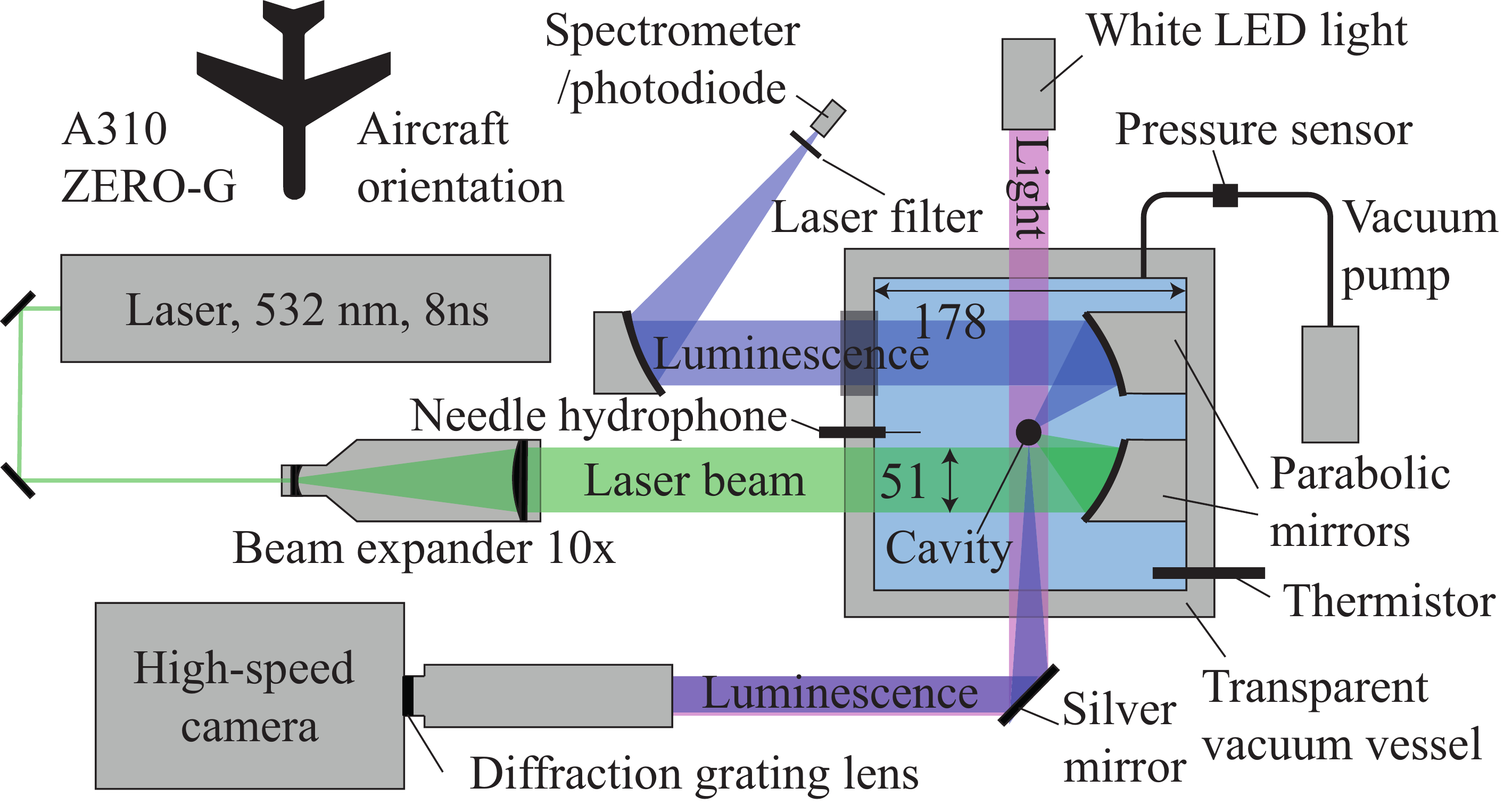}
\caption{(Color online) Schematic top-view of the experimental setup. The dimensions are given in mm.}
\label{fig:setup}
\end{center}
\end{figure}

Figure~\ref{fig:setup} shows a schematic of our experiment.
We generate highly spherical bubbles by using an immersed parabolic mirror to focus a Q-switched Nd:YAG laser (532~nm, 8~ns) in the middle of a cubic test chamber filled with demineralized water.
The water is initially partially degassed to remove large bubbles from the container boundaries, but we presume the water to be mostly air-saturated.
The bubbles are so spherical that the dominant pressure field anisotropy deforming the bubble is the gravity-induced pressure gradient~\cite{Obreschkow2013}.
Furthermore, owing to their high sphericity, these bubbles do not suffer a fission instability, i.e.~bubble splitting~\cite{Baghda1999,Baghda2001}, during their collapse, allowing very large bubbles to compress their enclosed gases efficiently and luminesce in the absence of external perturbations.
We obtain the bubble's maximum radius $R_{0}$ by measuring its collapse time $T_{c}$ (i.e.~half oscillation time) of the bubble with a needle hydrophone, which detects the passage of the shock waves emitted at the generation and the collapse of the bubble.
The maximum bubble radius is then obtained via $R_{0}=1.093T_{c}(\Delta p/\rho)^{1/2}$~\cite{Rayleigh1917}, where $\Delta p= p_{0}-p_{v}$ is the driving pressure ($p_{0}$ being the static pressure at the height of the bubble and $p_{v}$ the liquid vapor pressure) and $\rho$ is the liquid density.
It is considered unnecessary to correct this relation for the bubble's asphericity as the deformations in this work remain weak.
The temperature of the water is recorded with a thermistor and kept at room temperature (294.2$\pm 1$~K), and $p_{v}$ is computed for each bubble individually using the Antoine equation.
Simultaneous visualizations of luminescence, the radial evolution of the bubble and the shock wave emission are made with an ultra-high-speed CMOS camera (Shimadzu HPV-X2) filming at $10\times10^{6}$ frames/s (fps) with an exposure time of 50$\pm$10~ns and a back-light LED.
%

The time-averaged luminescence spectrum from a single bubble collapse is captured in the dark by a spectrometer (Ocean Optics QEPro, exposure time 8~ms).
The light emitted during the bubble collapse is collected using a second, aluminum-coated immersed parabolic mirror that reflects it through a fused silica-window (for UV transparency) onto another parabolic mirror outside the test chamber.
We chose aluminum-coated mirrors for their good UV reflection quality.
The external mirror focuses the light through a laser-blocking filter onto the entrance of the optical fiber that leads to the spectrometer. 
Without the filter the laser would saturate the measured spectrum despite the spectrometer being triggered only after the bubble generation.
%
\begin{figure}
\begin{center}
\hspace{1.3cm}\includegraphics[width=0.5\textwidth, trim=0cm 0cm 0cm 0cm, clip]{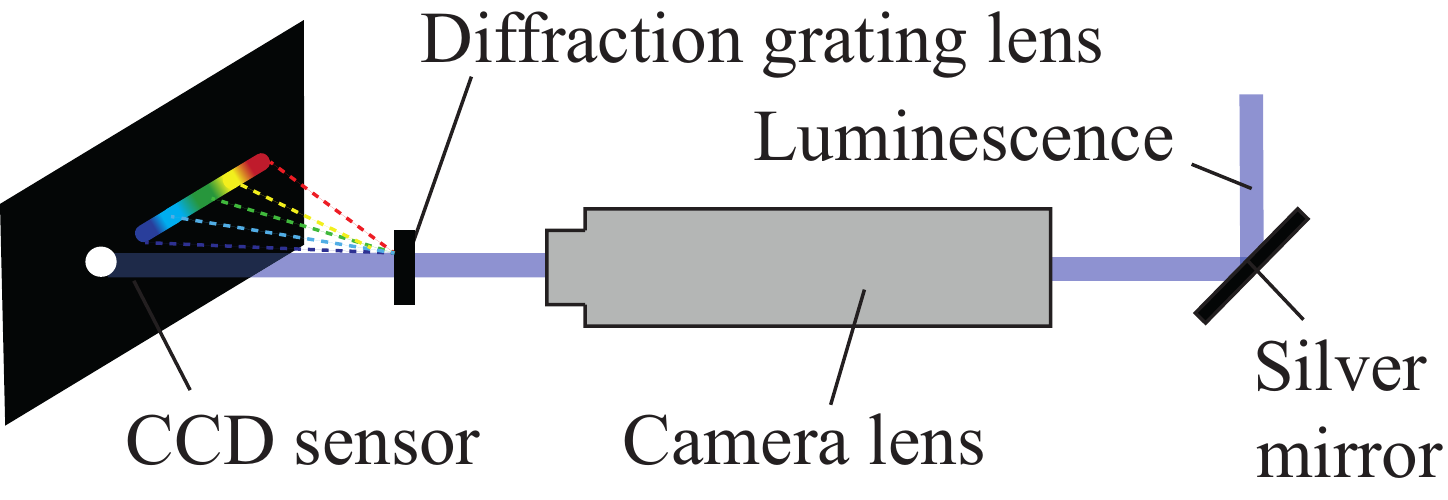}\\
\includegraphics[width=0.6\textwidth, trim=0cm 0cm 0.9cm 0cm, clip]{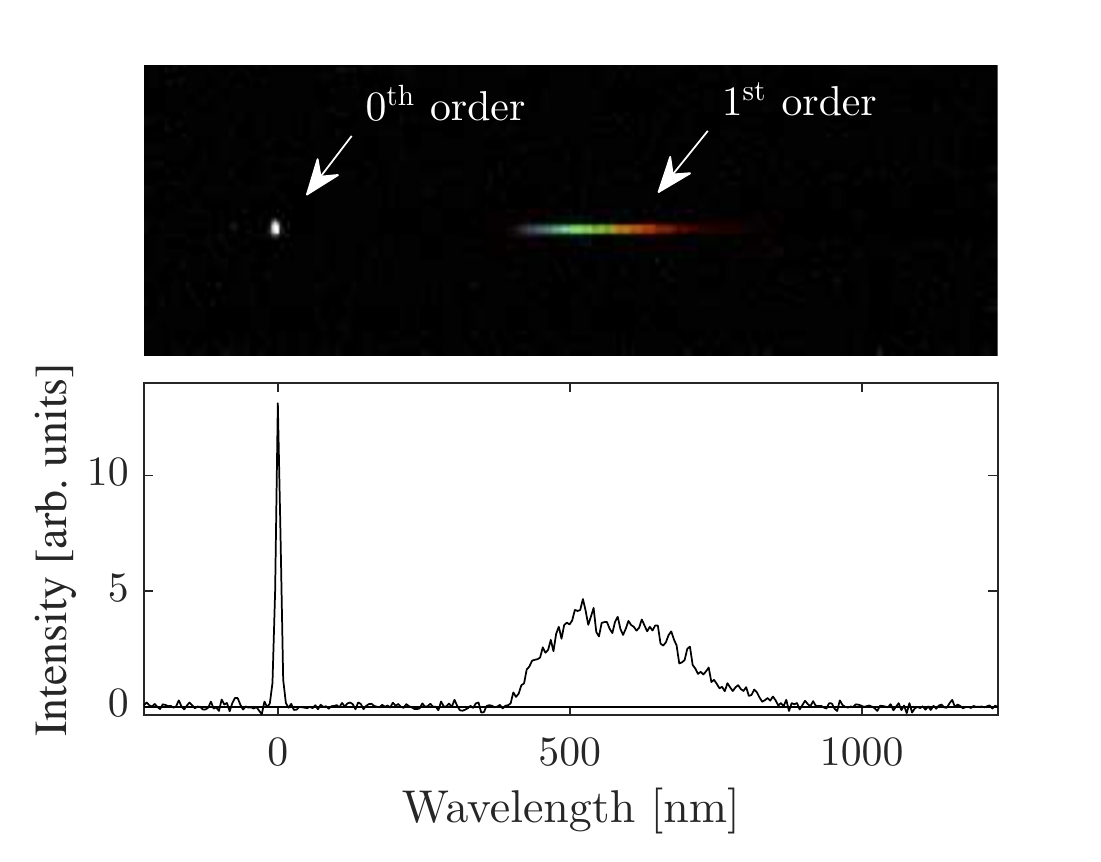}
\caption{(Color online) Schematic of the CCD luminescence detection system coupled with a diffraction grating lens (\textit{top}), a typical luminescence signal (0$^{\rm th}$ and 1$^{\rm st}$ order spectrum) as recorded by the CCD sensor (\textit{middle}) and the corresponding raw spectrum obtained from the pixel intensities of the image (\textit{bottom}).}
\label{fig:CCD}
\end{center}
\end{figure}

The luminescence spectrum is simultaneously measured by a second high-speed camera (Photron SA1.1) that has a CCD sensor (in place of the CMOS camera).
It is equipped with an astronomy-quality diffraction grating lens (RSpec, Star Analyzer SA-100) and films at $10^{5}$~fps with an exposure time of 10~$\mu$s. 
The reason for using the CCD over the CMOS camera to measure the spectrum is that it guarantees the luminescence to be fully contained in its exposure time, which the latter cannot.
The grating lens, placed between the camera objective and the CCD sensor, splits and deviates the light one or more diffraction orders located in a plane perpendicular to the grating lines, thus providing a spectrum on the sensor.
A schematic of the CCD light detection system is shown in Fig.~\ref{fig:CCD} along with a typical measured luminescence signal.
The reasons behind measuring the spectrum additionally with the camera are that it fills in the spectral gap in the spectrometer ($\sim500$--700~nm) caused by the laser-blocking filter and, more importantly, corrects the intensity of the spectrum recorded by the spectrometer, which is affected by the bubble's migration away from the parabolic mirror's focal point.
The bubble's displacement becomes important in particular at higher gravity levels for large bubbles that experience a strong Kelvin impulse~\cite{Supponen2016} (i.e.~the integrated momentum of the liquid during the growth and the collapse of the bubble~\cite{Blake1988}).
Such a displacement can weaken the signal measured by the spectrometer, and therefore it is corrected using the spectrum recorded by the CCD.
The CCD spectrum measurement is unaffected by the bubble's displacement as the luminescence spot stays within the image plane.

\begin{figure}
\begin{center}
\includegraphics[width=0.8\textwidth, trim=0cm 0cm 0cm 0cm, clip]{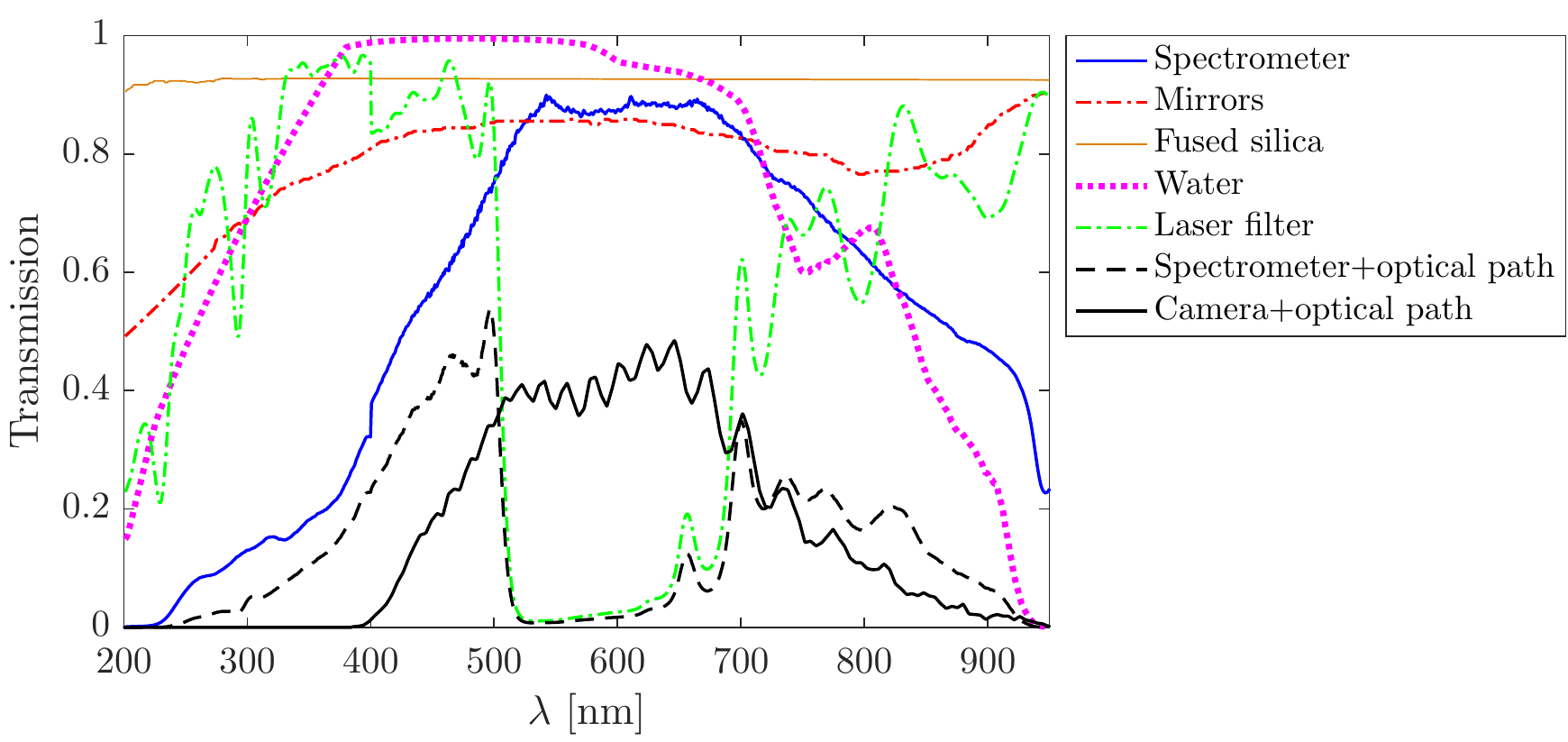}
\caption{(Color online) Transmission of light as a function of wavelength $\lambda$ for the various elements on the optical path from the luminescence emission point to the detectors.}
\label{fig:transmissions}
\end{center}
\end{figure}

The optical path from the luminescence to the spectrometer includes 194~mm water, 6~mm fused silica, two aluminum-coated parabolic mirrors and the laser filter.
To reach the camera's CCD sensor, the luminescent light travels through water, acrylic glass, a silver mirror, the camera lens and the grating lens.
The wavelength-dependent transmissions of the various elements in the optical paths are shown in Fig.~\ref{fig:transmissions}. 
The calibration of the spectrometer detector and the absorption/transmission spectra of the various optical components were provided by their respective manufacturers.
Water's absorption spectrum in the wavelength range of interest is found in the literature~\cite{Hale1973}.
The spectrum measured by the high-speed camera with the grating filter has been calibrated in-house combining the transfer functions of the camera and the optical path using a thermal light source placed inside the test chamber at the location where the bubble is generated.
This innovative luminescence measurement system allows for 1)~the collection of substantial amount of light from the rapid, small and weak luminescence of a single bubble collapse, 2)~the capturing of a wide spectrum from a single bubble collapse, including the UV, and 3)~staying far from the bubble not to disturb its dynamics.

Time-resolved measurements of the luminescence pulse are made using the same optical path as described above for the spectrometer, but by focusing the light onto a high-speed photodetector (Thorlabs, DET10A/M Si detector) without a laser-blocking filter.
The detector has a 1~ns rise time and is sensitive in the 200--1100~nm wavelength range.
The photodetector signal is recorded by an oscilloscope (4~GHz bandwidth), which is triggered using the hydrophone signal of the collapse shock wave and by applying a $25$~$\mu$s negative delay to account for the shock wave propagation over a distance of $\sim37$~mm to reach the hydrophone after the bubble collapse.

%
%

Three parameters influencing the bubble luminescence can be independently varied in our experiment: (i) the driving pressure $\Delta p\equiv p_{0}-p_{v}$ (0.06--1~bar), where $p_{0}$ is adjusted using a vacuum pump; (ii) the bubble energy $E_{0} = (4\pi/3) R_{0}^{3}\Delta p$ (0.4--28~mJ), adjusted by the laser pulse energy; and (iii) the constant, uniform pressure gradient $\boldsymbol{\nabla} p$ (=$\rho\mathbf{g}$, with the perceived gravitational acceleration $\left|\mathbf{g}\right|$ varied between 0--2~$g$, where $g=9.81$~ms$^{-2}$), modulated aboard European Space Agency parabolic flights (58th, 60th and 62nd parabolic flight campaigns) and on the first Swiss parabolic flight. 
The interest in using the hydrostatic pressure gradient to deform bubbles is its uniformity in space and time, in contrast to near boundaries.
This is an advantage in particular as it probes the influence of pressure gradients induced by any other inertial forces in addition to gravity.
Moreover, any practical instant of a smooth pressure field can be approximated to first order by such a uniform pressure gradient, thus extending the scope of this study to any situation involving bubbles in anisotropic pressure fields~\cite{Obreschkow2011,Supponen2016}.
These variables yield a wide range of maximum bubble radii, $R_{0}\sim1.5$--$10$~mm.
Such large bubbles present the advantage of easier resolution of the time and space scales associated with their collapse, in contrast to e.g.~SBSL experiments.
Additional details on the experiment and the parabolic flights may be found in Ref.~\cite{Obreschkow2013}.

We account for the effect of bubble asphericity due to the gravity-induced pressure gradient through the anisotropy parameter $\zeta\equiv \left|\boldsymbol{\nabla}p\right| R_{0}\Delta p^{-1}$, which is the dimensionless equivalent of the Kelvin impulse~\cite{Supponen2016,Obreschkow2011,Blake1988}.
$\zeta$ is varied by adjusting the maximum bubble radius $R_{0}$, the driving pressure $\Delta p$ and the pressure gradient $\left|\boldsymbol{\nabla} p\right|$ (through variable gravity).
Measuring at variable gravity allows for the decoupling of the roles of the driving pressure ($\Delta p$) and bubble deformation ($\zeta$), which is important because the expression of $\zeta$ for gravity-induced deformation includes $\Delta p$.
The pressure field anisotropy caused by the nearest boundary in our experiment is considered with $\zeta = -0.195\gamma^{-2}$ (which represents the dimensionless Kelvin impulse for bubbles near boundaries~\cite{Supponen2016}), where $\gamma$ is the stand-off parameter $\gamma=s/R_{0}$ and where $s=55$~mm is the distance between the bubble center and the parabolic mirror. 
The resultant $\zeta$ is given by the vector sum of the respective directional $\boldsymbol{\zeta}$.
We expect luminescence to vary with $\zeta$, since an increasing $\zeta$ implies stronger bubble deformation which, in turn, affects the different events associated with the bubble collapse, such as microjets~\cite{Supponen2016,Obreschkow2011} and shock waves~\cite{Vogel1988,Supponen2017}.

\section{Spectral analysis in variable gravity}
\label{s:spectra}

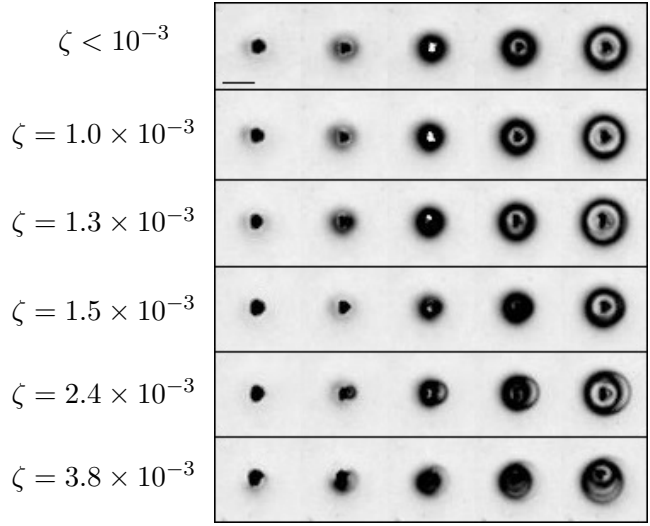
\begin{figure}
\begin{overpic}[width=0.35\textwidth]{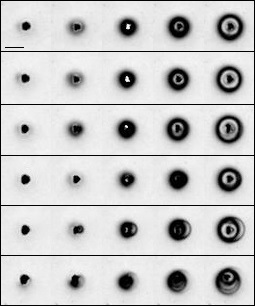}
\put (-30,91) {$\zeta < 10^{-3}$}
\put (-39,73) {$\zeta = 1.0\times 10^{-3}$}
\put (-39,56) {$\zeta = 1.3\times10^{-3}$}
\put (-39,39) {$\zeta = 1.5\times10^{-3}$}
\put (-39,23) {$\zeta = 2.4\times10^{-3}$}
\put (-39,7) {$\zeta = 3.8\times10^{-3}$}
\end{overpic}
\caption{Visualization of luminescence emitted at the final collapse stage of a single cavitation bubble at various $\zeta$. The luminescent flash is visible in the middle frame and followed by the rebound. The interframe time is 100~ns, the exposure time is 50~ns and the black bar shows the 1-mm scale. The bubble energy is the same in all cases ($E_{0}\approx 27$~mJ) and $\zeta$ varied by adjusting the driving pressure, from top to bottom, as $\Delta p=98$, 78, 58, 48, 28 and 18~kPa, yielding maximum bubble radii of $R_{0}=4.1$, 4.3, 4.8, 5.1, 6.1 and 7.1~mm. These bubbles were imaged on-ground at normal gravity.}%
\label{fig:lum}
\end{figure}

Selected images of high-speed movies visualizing luminescing bubbles of same energy $E_{0}$ collapsing at different levels of $\zeta$ at normal gravity are shown in Fig.~\ref{fig:lum}. 
The bubble interface, the luminescence and the sharp shock waves are captured in the same movie, owing to the short exposure time (50~ns). 
We observe a weakening of the luminescent flash for an increasing $\zeta$. 
One may also see a pronounced deflection of light near the bubble wall in the frames preceding the luminescence, which is due to the pressure rise in the surrounding liquid predicted by Lord Rayleigh a century ago~\cite{Rayleigh1917}.
At $\zeta = 3.8\times10^{-3}$ there is no visible luminescence and the bubble's deformation is clearly manifested by the emitted shock wave(s) no longer being spherically symmetric.

\begin{figure}
\begin{center}
\includegraphics[width=0.8\textwidth, trim=0cm 0cm 0cm 0cm, clip]{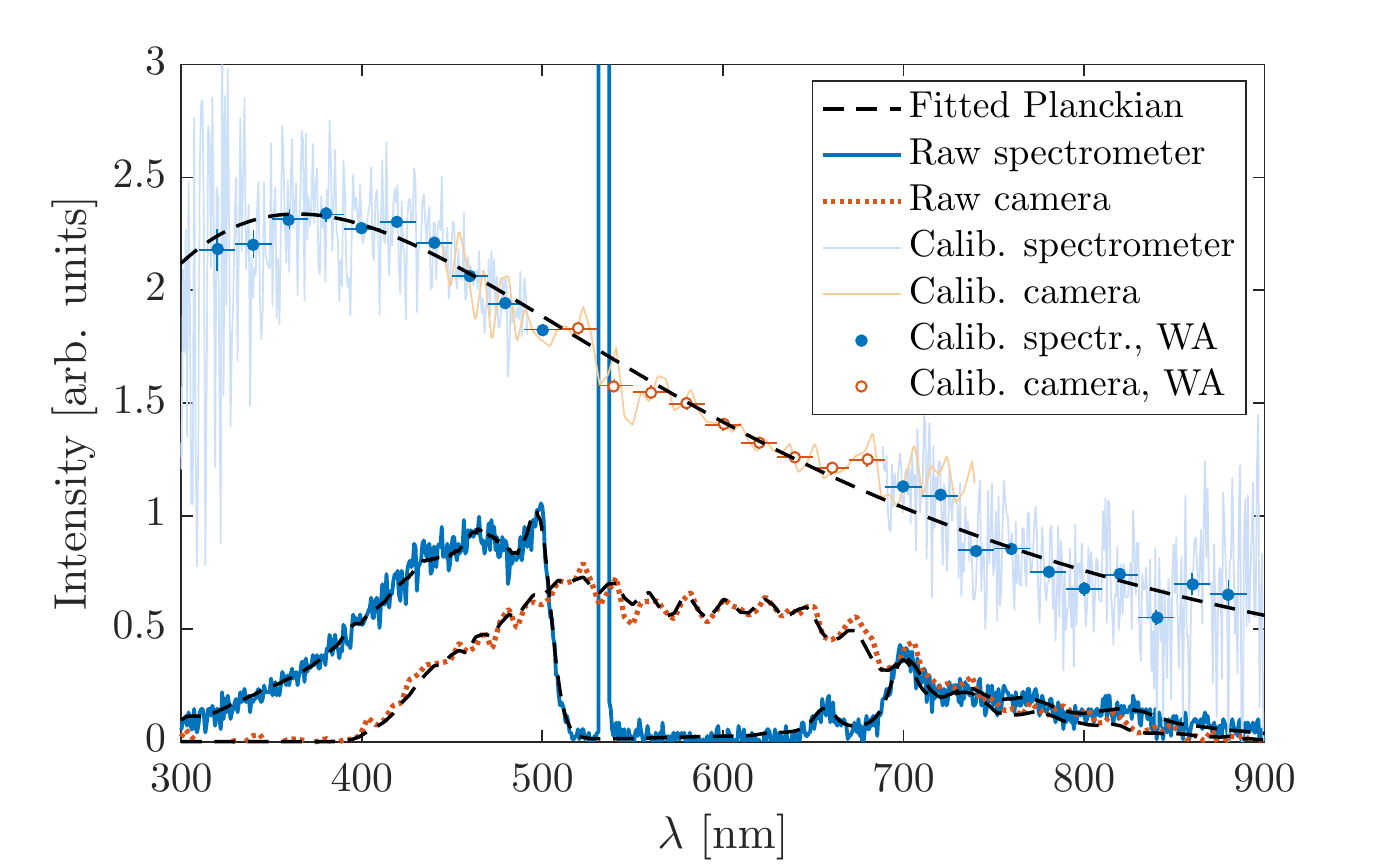}
\caption{(Color online) Typical luminescence spectrum from a single bubble collapse, measured with the spectrometer with an exposure time 8~ms and the high-speed CCD camera with an exposure time of 10~$\mu$s. Both raw and calibrated spectra are shown, together with the fitted Planckians. Window averages (WA) of 20~nm-large windows are also displayed. The peak around 532~nm is caused by the strong laser pulse despite the $>99\%$ attenuation of the filter. Here $R_{0}=3.0$~mm, $\Delta p=78$~kPa and $\left|\mathbf{g}\right|=1$~$g$.}%
\label{fig:spec}
\end{center}
\end{figure}

The luminescence spectrum is well approximated by the blackbody model~\cite{Baghda2001,Hiller1992,Lepoint1996}, and since the bubble temperature cannot be directly measured, a fitted blackbody provides a reasonable estimation for it. 
The effective blackbody temperature and energy of luminescence can be inferred by fitting the spectra with a Planckian function of the form: 
\begin{equation}
L(\lambda,I,T_{\rm lum}) = A\frac{I}{\lambda^{5}}\frac{1}{\exp\left(\frac{hc}{\lambda k_{B}T_{\rm lum}}\right)-1}\ [\text{J/nm}]
\label{eq:black}
\end{equation}
where $\lambda$ is the wavelength, $h$ and $k_{B}$ are the Planck and Boltzmann constants respectively, $c$ is the speed of light, $A$ is a constant prefactor determined from calibration, $T_{\rm lum}$ is the blackbody temperature and $I$ stands for the product of the luminescence pulse duration and the projected emitting surface (which cannot be disentangled with the spatial and temporal resolution of our apparatus). 
The best-fit values are obtained by fitting Eq.~(\ref{eq:black}), after correcting it for the absorption losses of Fig.~\ref{fig:transmissions}, with the measured raw spectra through maximum likelihood for the pair ($E_{\rm lum}$,$T_{\rm lum}$), where $E_{\rm lum} = IT_{\rm lum}^4$ is the luminescence energy through Stefan-Boltzmann law.
The estimated standard error of the maximum likelihood fit is obtained from the covariance matrix (estimated via the inverse of the Hessian matrix) representing the goodness of fit to the data.
Figure~\ref{fig:spec} displays a typical measured luminescence spectrum from a single bubble collapse.
%

We estimate the total luminescence energy $E_{\rm lum}$ by assuming a uniform light emission in the solid angle of 4$\pi$.
In this way, 6.7\% of all the photons are expected to reach the calibrated spectrometer detector.
We use as a reference a highly spherical bubble collapsing in microgravity, that is assumed to undergo no displacement from the focal point of the parabolic mirror.

\begin{figure}
\begin{center}
\begin{overpic}[width=0.48\textwidth, trim=0cm 0cm 0cm 0cm, clip]{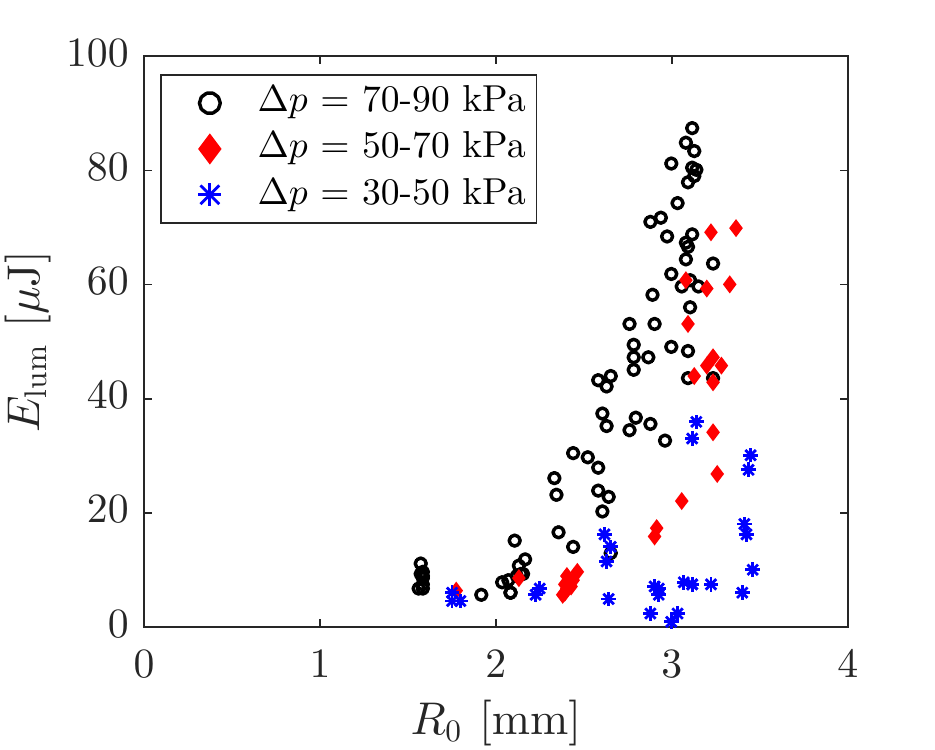}
\put (17,49) {(a)}
\end{overpic}
\begin{overpic}[width=0.48\textwidth, trim=0cm 0cm 0cm 0cm, clip]{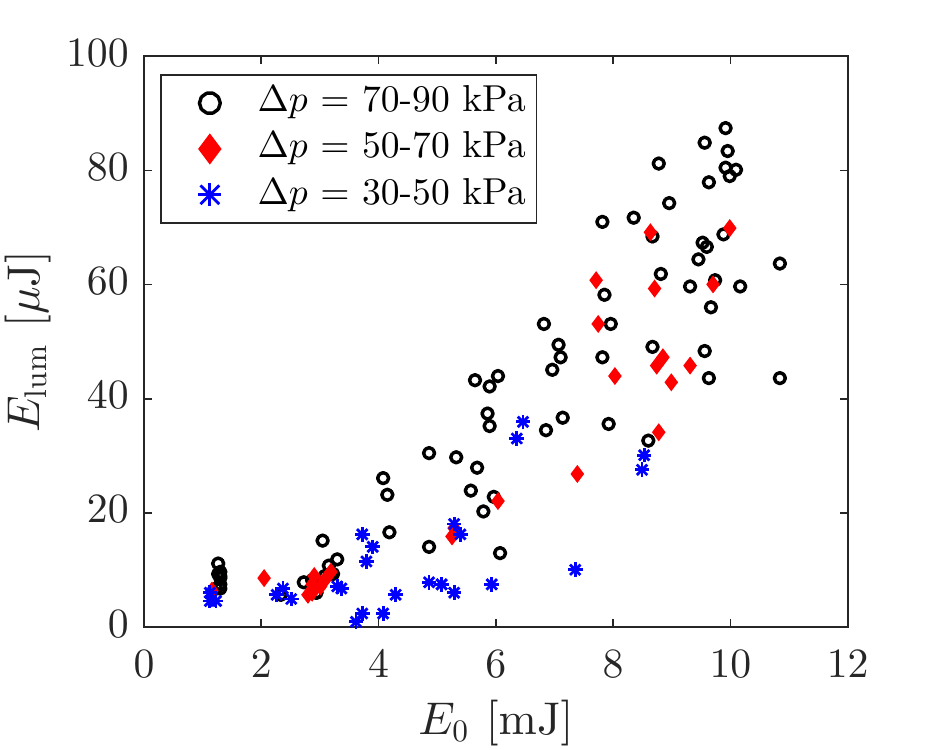}
\put (17,49) {(b)}
\end{overpic}
\caption{(Color online) Luminescence energy $E_{\rm lum}$ as a function of (a)~maximum bubble radius $R_{0}$ and (b)~bubble energy $E_{0}$ for three ranges of driving pressures $\Delta p$. Each point corresponds to a measurement from a single, spherical collapse ($\zeta < 7\times10^{-4}$).}%
\label{fig:R_Elum}
\end{center}
\end{figure}

Figure~\ref{fig:R_Elum}(a) shows the luminescence energy $E_{\rm lum}$, obtained through the best Planckian fit, as a function of the maximum bubble radius $R_{0}$ for three different ranges of driving pressure $\Delta p$.
Only bubbles collapsing highly spherically ($\zeta<7\times10^{-4}$) have been selected in order to exclude deformation-induced hindering of the luminescence, and the data include bubbles collapsing in microgravity.
The maximum radii are within the range $R_{0}=1.5$--$3.5$~mm, which, to our knowledge, extend to the largest reported laser-induced luminescing bubbles collapsing freely and spherically in water. 
As expected, one may observe an increase of $E_{\rm lum}$ with growing $R_{0}$ for a fixed $\Delta p$, the tendency being consistent with the literature~\cite{Ohl1998,Baghda1999,Ohl2002}.
In the literature, however, a decrease of luminescence energy for laser-induced bubbles with increasing maximum radii beyond $R_{0}\approx 1.5$~mm has also been reported~\cite{Ohl2002}.
This is likely attributed to the use of less point-like focusing methods (e.g.~converging lens) that yield bubbles that are more disturbed in the collapse phase and cause e.g.~bubble splitting~\cite{Baghda1999,Brujan2005}, such disturbances being enhanced for increasing bubble radius.
Bubbles with $R_{0}>3.5$~mm in our experiment are affected by the nearest surface, i.e.~the parabolic mirror at a distance of 55~mm from the bubble center, which is accounted for in $\zeta$.
%

For a given $R_{0}$, a lower $\Delta p$ yields weaker luminescence, which is expected since $E_{\rm lum} \propto E_{0}=(4\pi/3) R_{0}^{3}\Delta p$~\cite{Wolfrum2001,Brujan2005}.
Figure~\ref{fig:R_Elum}(b) verifies this relation, but still suggests slightly weaker luminescence energies for bubbles collapsing with a lower $\Delta p$.
This result is consistent with the past observation of more energetic luminescence from bubbles collapsing at higher static pressures for a fixed $E_{0}$~\cite{Wolfrum2001}.
Bubbles at a low $\Delta p$ have a longer collapse time and thereby an increased surface area and interaction time, possibly yielding increased energy loss by thermal conduction or mass flow by nonequilibrium evaporation/condensation at bubble wall~\cite{Wolfrum2001}.

The important scatter of our results is due to the limited reproducibility of the luminescence. 
We find the spectral intensities between individual bubbles in the same conditions to vary by approximately 45\%, while the maximum bubble radii vary by less than 1\%.
These brightness fluctuations are likely related to the microscopic size of the luminescent plasma, which makes it highly sensitive to minor perturbations and easily obscured by nuclei and impurities in the water. 
%
%
\begin{figure}
\begin{center}
\includegraphics[width=.5\textwidth, trim=0cm 0cm 0cm 0cm, clip]{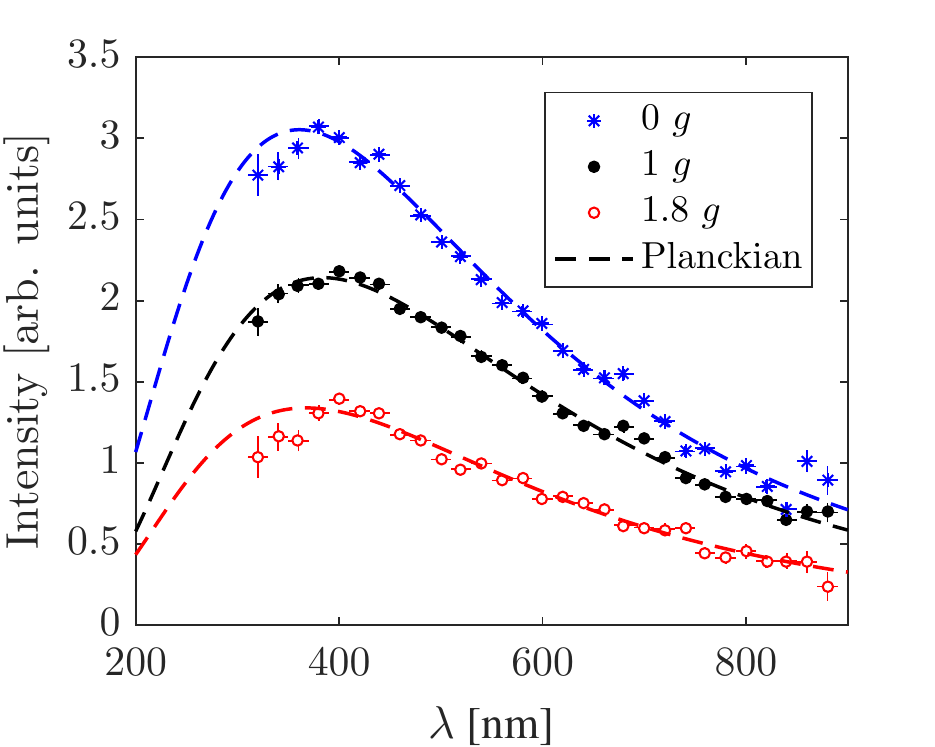}
\caption{(Color online) Single cavitation bubble luminescence spectra at three different gravity levels for the same laser pulse energy ($R_{0}=3\pm0.1$~mm) and static pressure of the water ($p_{0}=81\pm1$~kPa). Each spectrum is measured at a single bubble collapse.}%
\label{fig:sp_g}
\end{center}
\end{figure}

Figure~\ref{fig:sp_g} displays three examples of typical spectra of single cavitation bubble luminescence with the only varying parameter being the perceived gravity level (0~$g$, 1~$g$ and 1.8~$g$). 
It is evident that the gravity-induced pressure gradient quenches the SCBL energy. 
Surprisingly, on none of the raw spectra do we observe a prominent peak corresponding to OH$^{-}$ or other emission lines at any wavelength, even for the most deformed luminescing bubbles.
This could, however, be due to the limited wavelength-resolution of our apparatus.
%
\begin{figure}
\begin{center}
\begin{overpic}[width=0.7\textwidth, trim=4.4cm 8cm 4.2cm 8cm, clip]{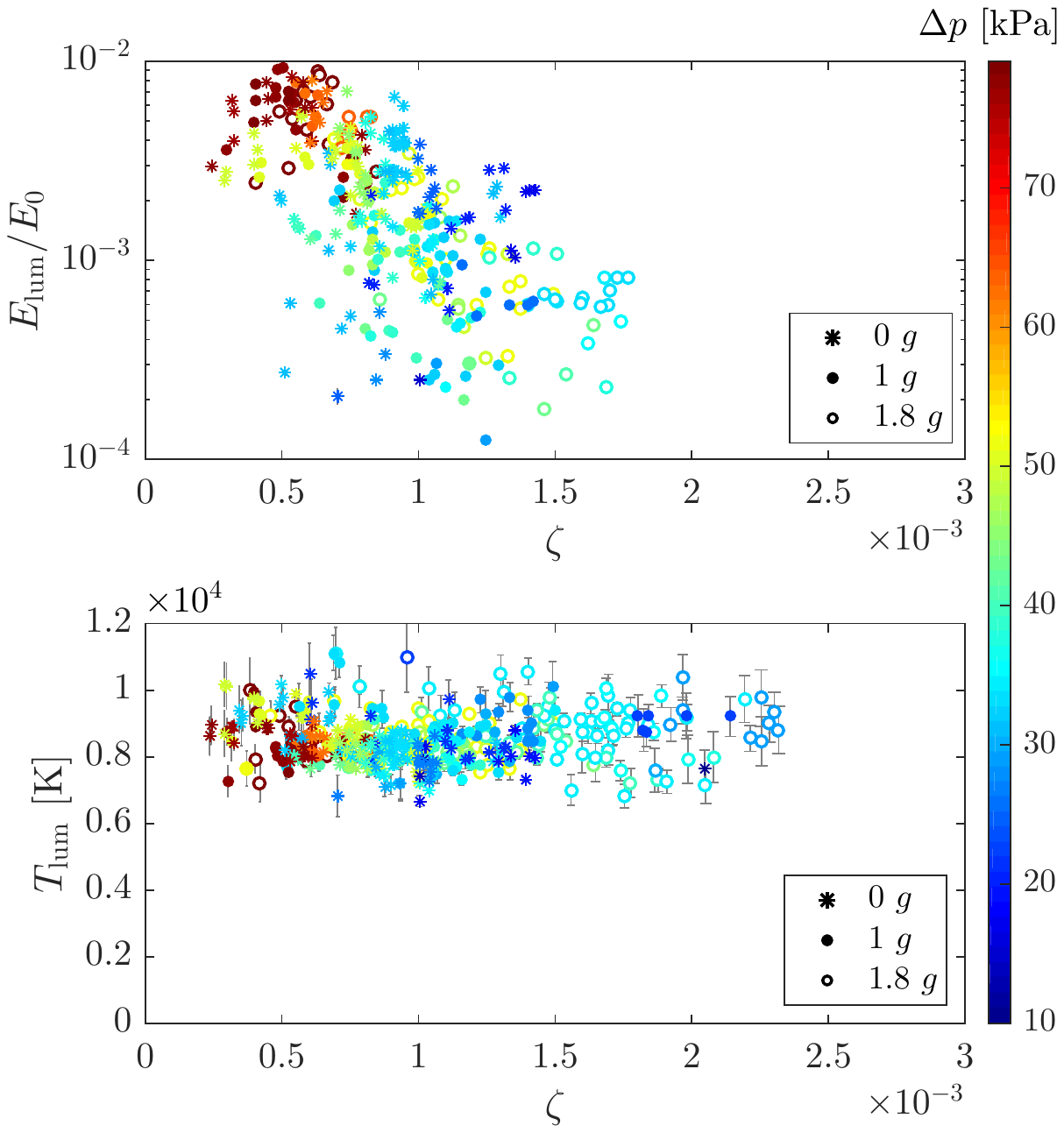}
\put (13,61) {(a)}
\put (13,13) {(b)}
\end{overpic}
\caption{(Color online) Single cavitation bubble luminescence (a)~relative energy $E_{\rm lum}/E_{0}$ and (b)~blackbody temperature $T_{\rm lum}$ as a function anisotropy parameter $\zeta$. Each data point represents a single bubble measurement. Colors (gray shades) indicate the driving pressures and symbols indicate different levels of gravity. The error bars indicate the $\pm \sigma$ uncertainty of the best-fit estimate of the blackbody temperature, while the error for the $E_{\rm lum}/E_{0}$ estimate is small ($\sigma\sim10^{-5}$).}%
\label{fig:zE_p}
\end{center}
\end{figure}

To quantify the fraction of bubble energy dissipated into luminescence, we normalize the luminescence energy $E_{\text{lum}}$ to the bubble energy $E_{0}$. 
We only retain the cases where luminescence is detected by both the spectrometer and the CCD camera.
Note that the CCD signal helps correcting the spectrum of the spectrometer if the bubble moves out of the focal point of the parabolic mirror during its collapse. 
The dependence of the relative luminescence energy on the anisotropy parameter $\zeta$ is displayed in Fig.~\ref{fig:zE_p}(a).
Here $\zeta$ is altered by a wide range of $R_{0}$, $\mathbf{g}$ and $\Delta p$ in order to disentangle their respective effects on luminescence from that of the bubble deformation.
The maximum $\Delta p$ was achieved when the test vessel reached the aircraft cabin pressure, i.e.~$p_{0}\approx80$~kPa.
The results show a rapid quenching of relative luminescence energy with increasing $\zeta$. 
Luminescence takes up to approximately 1\% of the bubble's initial energy.
The rest of the bubble's energy is distributed predominantly into the shock wave emission and the formation of a rebound bubble for spherically collapsing bubbles~\cite{Tinguely2012}.
Owing to microgravity, we are able to create large bubbles, which in normal gravity would be deformed, that collapse highly spherically at low $\Delta p$ and emit luminescence.
Correspondingly, higher gravity levels allow us to stretch the range of $\zeta$ to higher values for a given $\Delta p$.
Up to the scatter, the data points exhibit a linear trend on a logarithmic scale as a function of $\zeta$ regardless of the gravity level. 
Luminescence is not detected by the spectrometer for anisotropy levels beyond $\zeta \approx 3.5\times10^{-3}$, which corresponds to the same Kelvin impulse at $\gamma \approx 7.5$ for bubbles deformed by neighboring surfaces~\cite{Supponen2016}.
Note that we only obtain reliable fitted blackbody energies, which require the CCD signal, up to $\zeta \approx1.8\times10^{-3}$ (in Fig.~\ref{fig:zE_p}(a)), due to the poor signal-to-noise ratio of luminescence from more deformed bubbles.

Figure~\ref{fig:zE_p}(b) displays our best-fit estimates of the bubble's blackbody temperatures as a function of $\zeta$. 
We obtain reliable fitted blackbody temperatures, which only require the spectrometer signal, up to $\zeta \approx 2.5\times10^{-3}$.
The temperatures fall in the range between $T_{\rm lum}=7000$--11500~K which is in good agreement with previous laser-induced bubble luminescence studies, in which the temperatures from averaged spectra varied between 7680~K (close to a solid surface) and 9150~K (at elevated ambient pressure)~\cite{Brujan2005,Brujan2005b}. 
This range, however, is attributed to the important scatter (which is expected owing to the experimental and fitting errors) rather than a clear relationship with the governing parameters.
The highly spherical bubbles with the highest luminescence energies do not exhibit higher blackbody temperatures than the luminescing deformed bubbles.
This result is in disagreement with the observations of Brujan and Williams~\cite{Brujan2005} who found the temperatures (estimated from averaged spectra) to decrease with decreasing distance between the bubble and a rigid boundary, that is, with increasing bubble deformation.

\section{Time-resolved measurements}

The luminescence pulse duration for spherically collapsing laser-induced cavitation bubbles has been shown to be in the nanosecond scale and to scale with the maximum bubble radius $R_{0}$.
For example, for $R_{0}=0.3$~mm, the full width at half maximum (FWHM) has been measured as $\tau\approx3$~ns~\cite{Baghda1999,Brujan2005}, for $R_{0}=1$~mm, $\tau\approx6$--$8$~ns~\cite{Ohl2002,Brujan2005} and for $R_{0}=1.8$~mm, $\tau\approx10$~ns~\cite{Ohl2002}.
Centimetric bubbles generated by a spark or expanded through a chemical reaction may luminesce for tens of microseconds~\cite{Zhang2017,Duplat2015}.
Owing to the high sphericity of the initial plasma generating the bubble, large bubbles in our experiment ($R_{0}>2$~mm) are able to collapse spherically without bubble splitting decreasing the efficiency of the final gas compression.
We therefore expect the luminescence pulse durations here to exceed those reported in the literature for laser-induced bubbles.

\begin{figure}
\begin{center}
\begin{overpic}[width=0.31\textwidth, trim=0cm 0cm 0cm 0cm, clip]{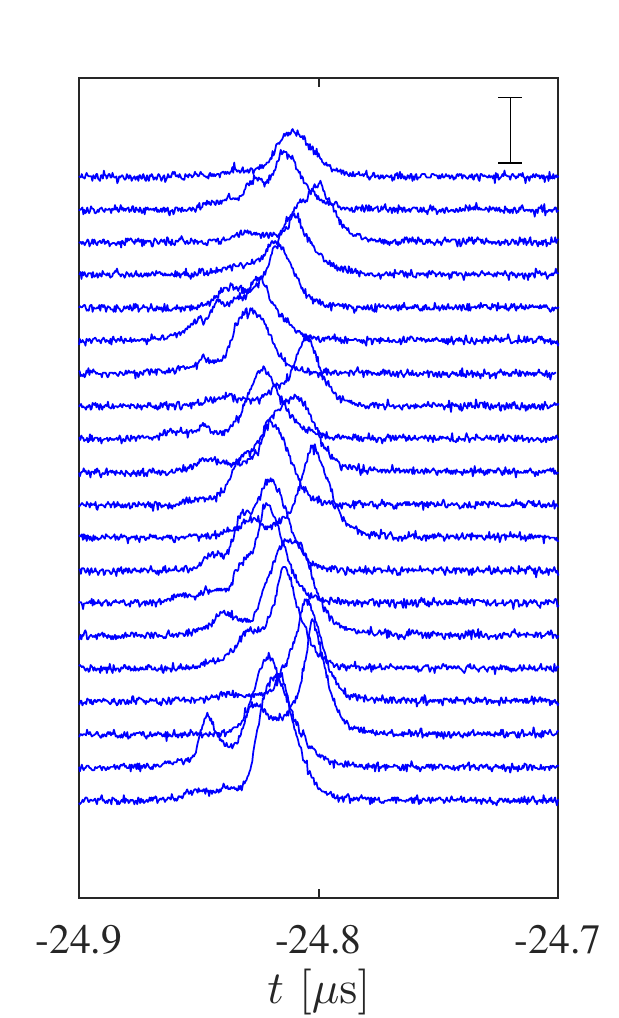}
\put (9,87) {(a)}
\put (35,85) {10 mV}
\put (9,14) {$\Delta p = 98$~kPa}
\end{overpic}
\begin{overpic}[width=0.31\textwidth, trim=0cm 0cm 0cm 0cm, clip]{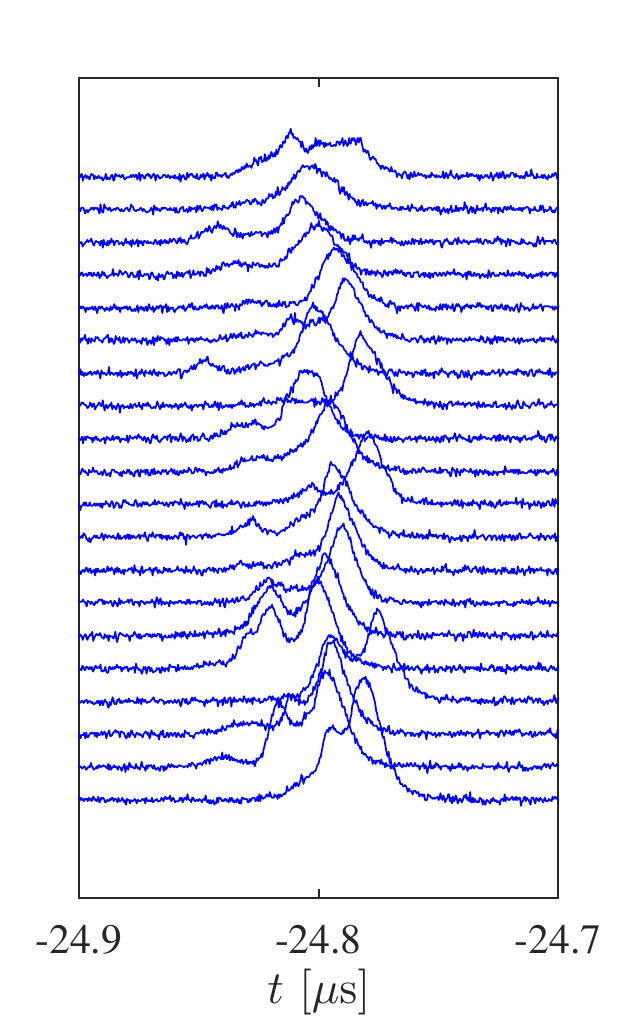}
\put (9,87) {(b)}
\put (9,14) {$\Delta p = 78$~kPa}
\end{overpic}
\begin{overpic}[width=0.31\textwidth, trim=0cm 0cm 0cm 0cm, clip]{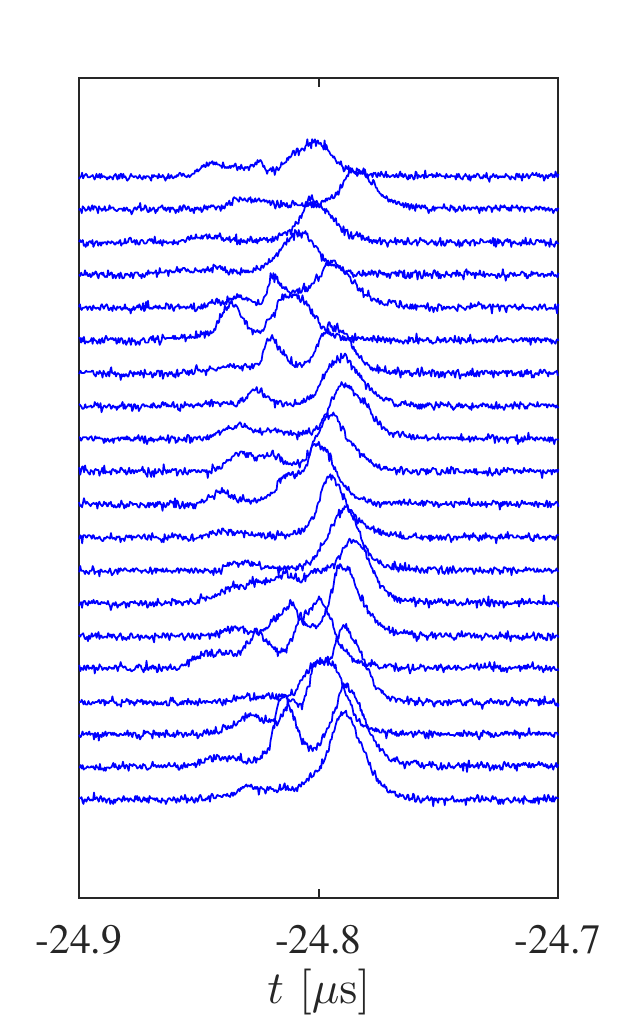}
\put (9,87) {(c)}
\put (9,14) {$\Delta p = 68$~kPa}
\end{overpic}
\begin{overpic}[width=0.31\textwidth, trim=0cm 0cm 0cm 0cm, clip]{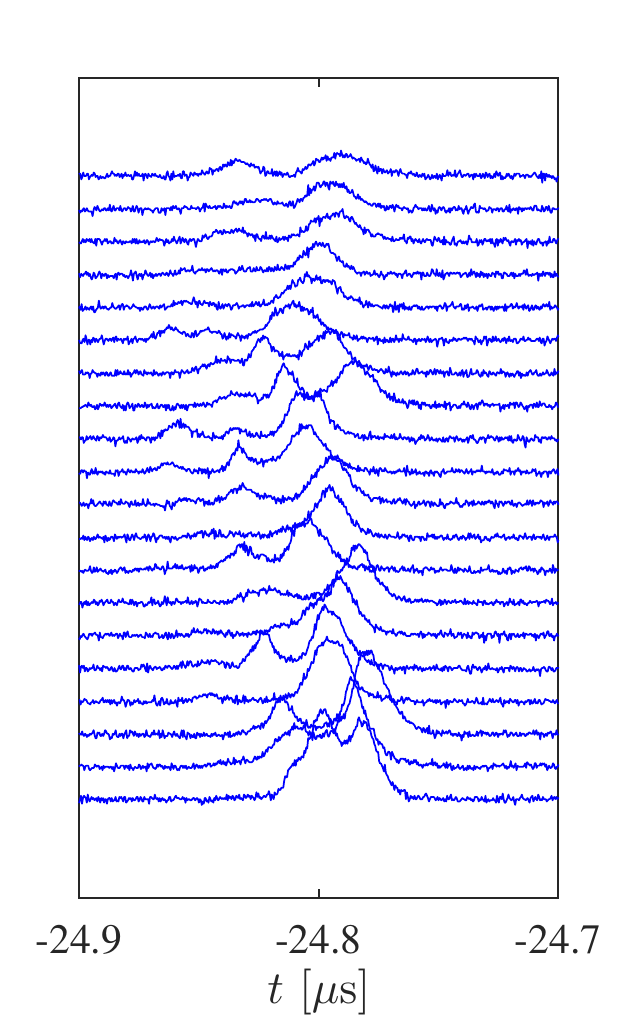}
\put (9,87) {(d)}
\put (9,14) {$\Delta p = 58$~kPa}
\end{overpic}
\begin{overpic}[width=0.31\textwidth, trim=0cm 0cm 0cm 0cm, clip]{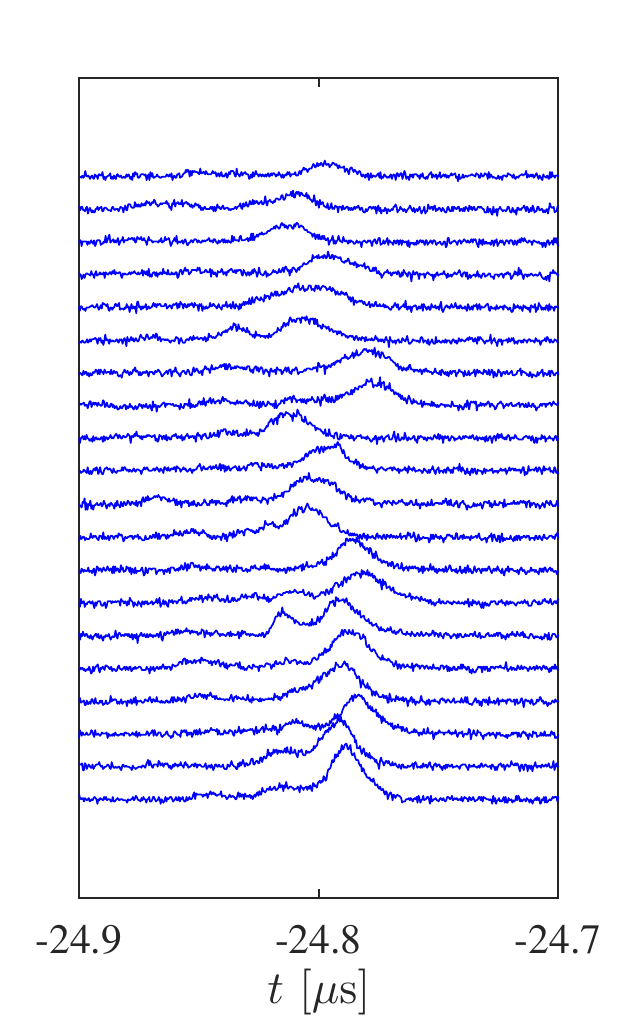}
\put (9,87) {(e)}
\put (9,14) {$\Delta p = 48$~kPa}
\end{overpic}
\begin{overpic}[width=0.31\textwidth, trim=0cm 0cm 0cm 0cm, clip]{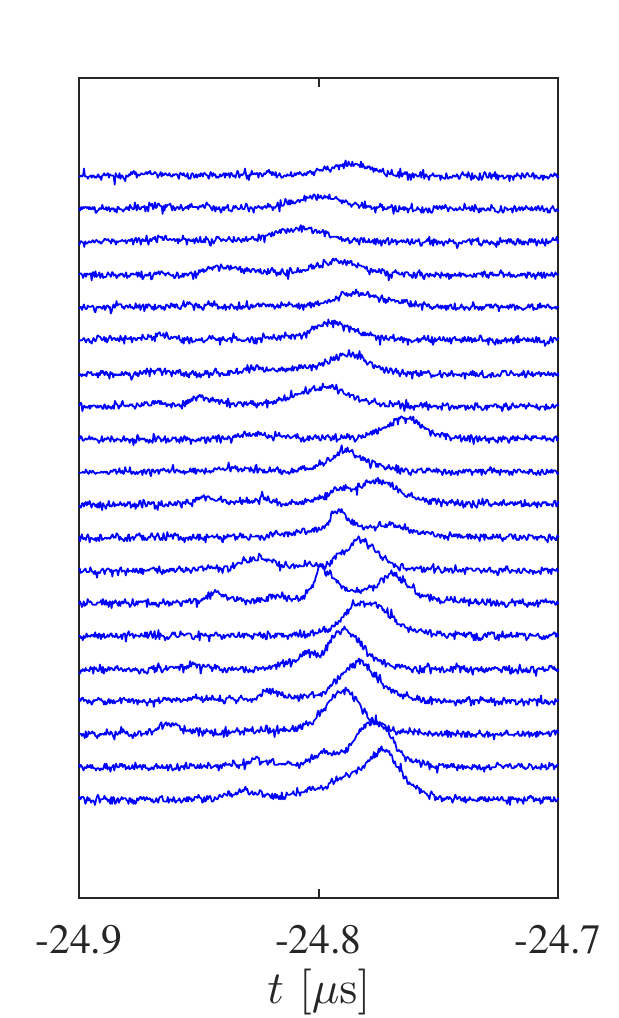}
\put (9,87) {(f)}
\put (9,14) {$\Delta p = 38$~kPa}
\end{overpic}
\caption{(Color online) Waterfall plots of the luminescence signals measured by the photodetector for different driving pressures (a)~$\Delta p=98$~kPa ($R_{0}=3.8$~mm, $\zeta=7.8\times10^{-4}$), (b)~$\Delta p=78$~kPa ($R_{0}= 4.0$~mm, $\zeta= 9.0\times10^{-4}$), (c)~$\Delta p=68$~kPa ($R_{0}= 4.2$~mm, $\zeta= 9.9\times10^{-4}$), (d)~$\Delta p=58$~kPa ($R_{0}= 4.5$~mm, $\zeta= 1.1\times10^{-3}$), (e)~$\Delta p=48$~kPa ($R_{0}= 4.7$~mm, $\zeta=1.3 \times10^{-3}$) and (f)~$\Delta p=38$~kPa ($R_{0}= 5.1$~mm, $\zeta= 1.6\times10^{-3}$). $E_{0} \approx 22$~mJ. Each plot contains 20 signals. The scaling shown in (a) is the same in all plots. $t=0$~$\mu$s corresponds to the instant at which the hydrophone detects the collapse shock. The standard deviations for $R_{0}$ and $\zeta$ are $\sigma_{R_{0}}\approx0.03$~mm and $\sigma_{\zeta}\approx1.5\times10^{-5}$, respectively. The measurements were made at normal gravity.}
\label{fig:water1}
\end{center}
\end{figure}
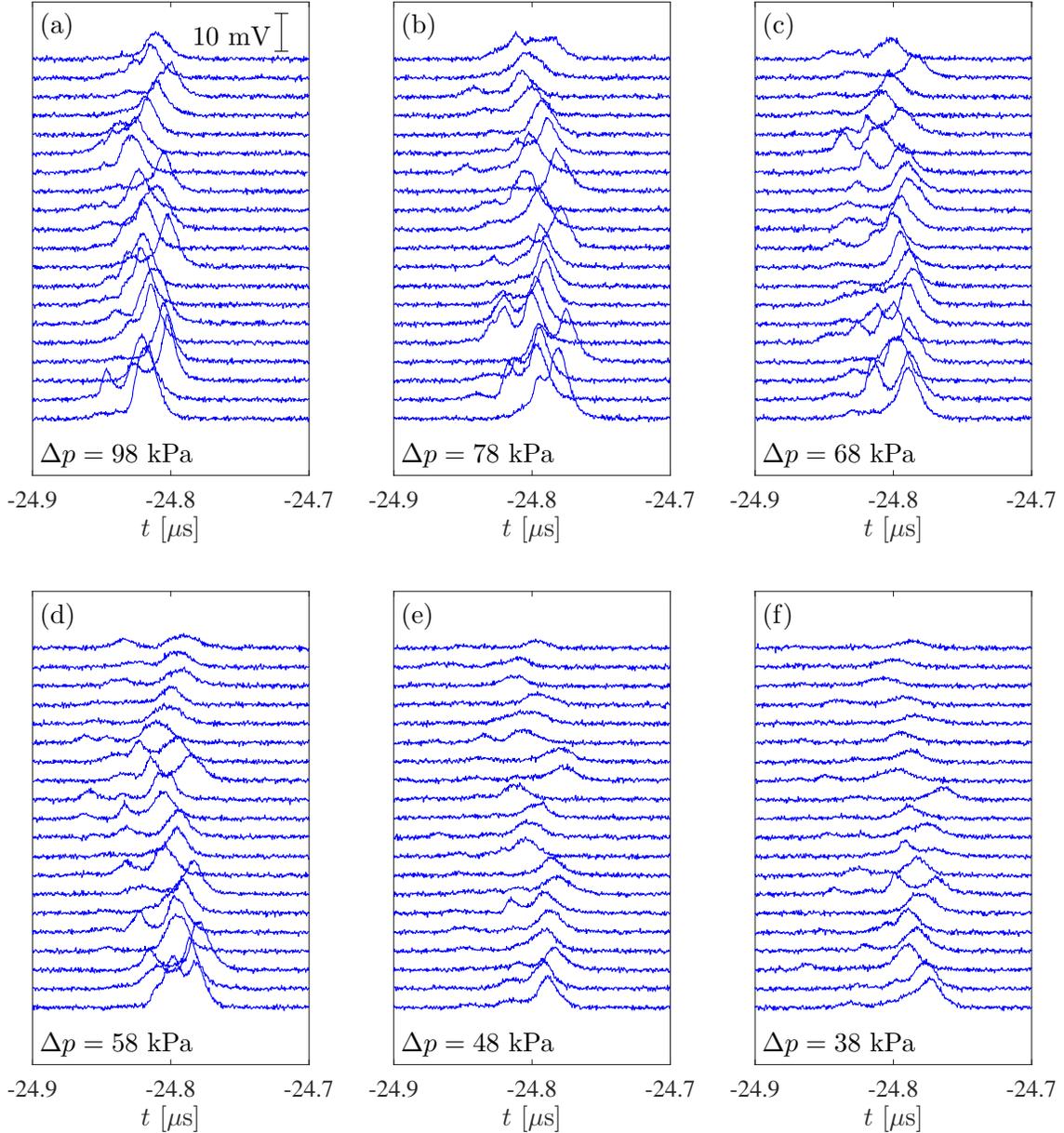
Figures~\ref{fig:water1}(a)--\ref{fig:water1}(f) show waterfall plots of 20 photodetector signals measured from single bubble collapses with a fixed bubble energy $E_{0}\approx22$~mJ and at different driving pressures $\Delta p$.
The signals are sorted such as the peak amplitudes are in descending order from bottom to top.
Here $t = 0$~$\mu$s corresponds to the moment at which the hydrophone detects the collapse shock, which has propagated a distance of 37~mm from the bubble.
It should be noted that the amplitudes of the photodetector signals are not corrected for the bubble displacement.
All photodetector measurements are made on-ground at normal gravity.
The standard deviation of the maximum peak timing with respect to $t=0$~$\mu$s ranges from $8$ to $12$~ns.
Consistent with the spectral analysis of Sec.~\ref{s:spectra}, the energy of the luminescence signals decreases with increasing $\zeta$.
The number of peaks in the photodetector signals varies between 1 and 4, suggesting multiple events yielding light emission.
Similar peaks have been observed in the past in photomultiplier tube measurements for both single and multiple bubble collapses~\cite{Ohl2002,Sukovich2012}.
Such multiple peaks are often randomly distributed in time with respect to the strongest peak, which, for the majority of cases, is the last event.
Figures~\ref{fig:water1}(a)--\ref{fig:water1}(c) show signals with up to two peaks and at lower driving pressures (Figs.~\ref{fig:water1}(d) and~\ref{fig:water1}(f)), where the amplitudes have substantially decreased, even three or four peaks may be observed.
The luminescence events occur within a time frame of approximately 200~ns.

\begin{figure}
\begin{center}
\begin{overpic}[width=.32\textwidth, trim=0cm 0cm 0cm 0cm, clip]{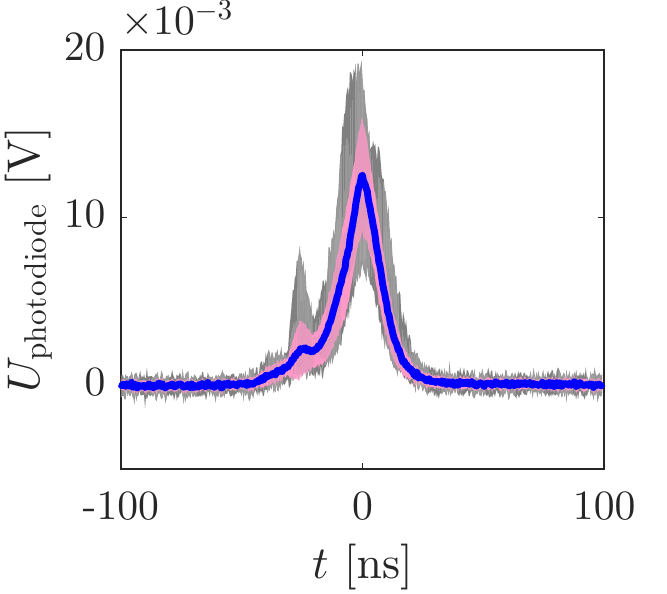}
\put (19,74) {(a)}
\put (58,74) {\footnotesize $\Delta p$=98kPa}
\put (57,66) {\footnotesize $R_{0}$=3.8mm}
\put (42,22) {$\zeta = 7.8\times10^{-4}$}
\end{overpic}
\begin{overpic}[width=.32\textwidth, trim=0cm 0cm 0cm 0cm, clip]{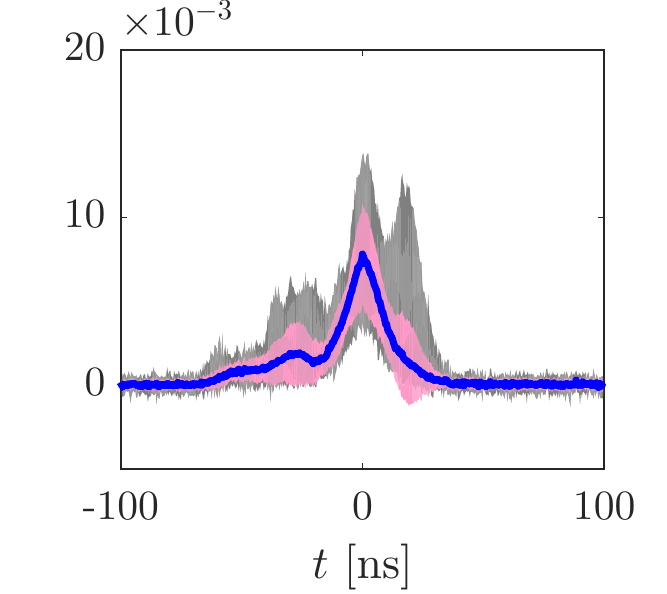}
\put (19,74) {(b)}
\put (58,74) {\footnotesize $\Delta p$=58kPa}
\put (57,66) {\footnotesize $R_{0}$=4.5mm}
\put (42,22) {$\zeta = 1.1\times10^{-3}$}
\end{overpic}
\begin{overpic}[width=.32\textwidth, trim=0cm 0cm 0cm 0cm, clip]{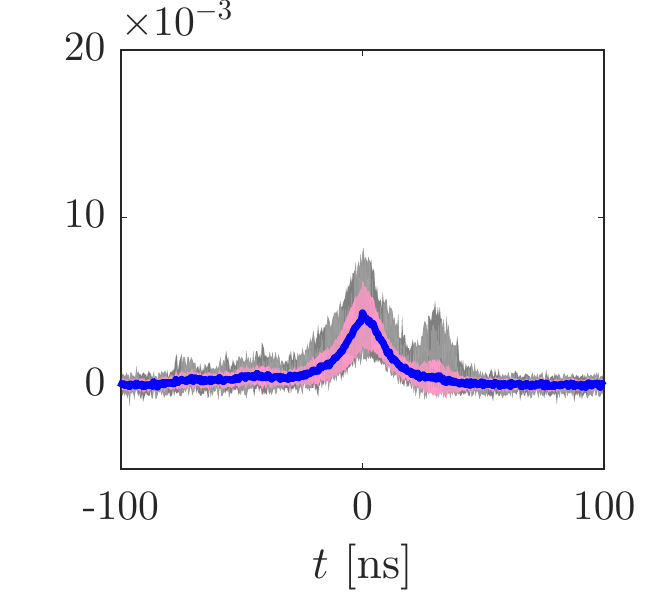}
\put (19,74) {(c)}
\put (58,74) {\footnotesize $\Delta p$=38kPa}
\put (57,66) {\footnotesize $R_{0}$=5.1mm}
\put (42,22) {$\zeta = 1.6\times10^{-3}$}
\end{overpic}
\begin{overpic}[width=.32\textwidth, trim=0cm 0cm 0cm 0cm, clip]{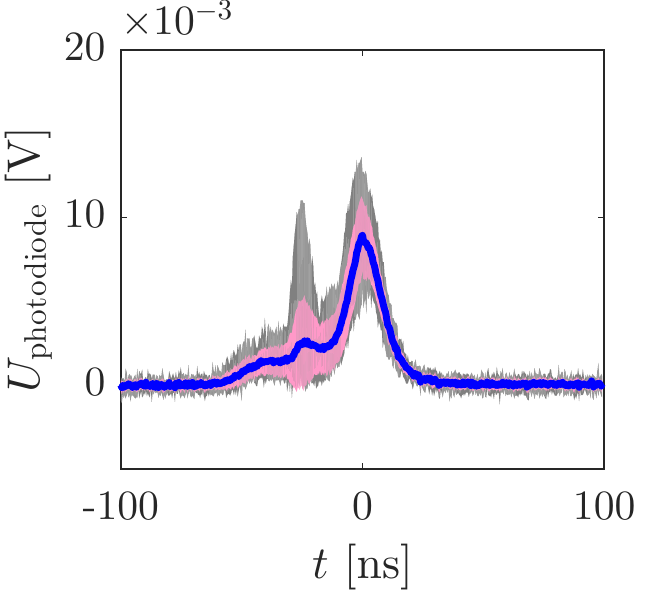}
\put (19,74) {(d)}
\put (58,74) {\footnotesize $\Delta p$=98kPa}
\put (57,66) {\footnotesize $R_{0}$=3.4mm}
\put (42,22) {$\zeta = 6.4\times10^{-4}$}
\end{overpic}
\begin{overpic}[width=.32\textwidth, trim=0cm 0cm 0cm 0cm, clip]{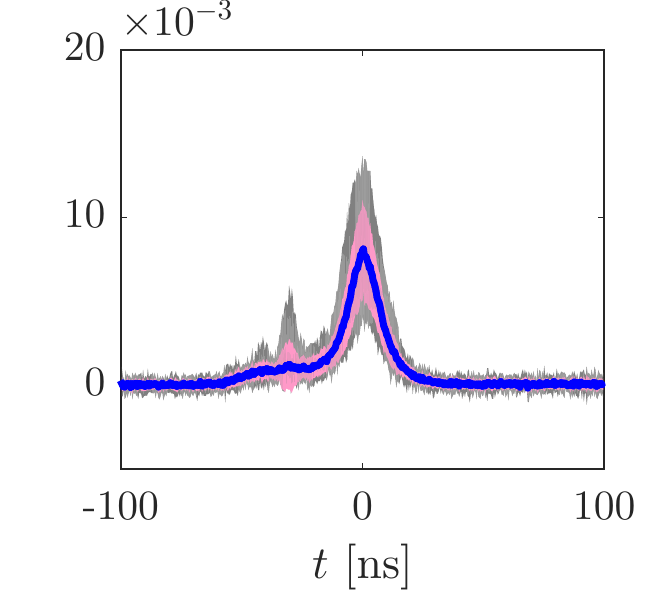}
\put (19,74) {(e)}
\put (58,74) {\footnotesize $\Delta p$=58kPa}
\put (57,66) {\footnotesize $R_{0}$=4.0mm}
\put (42,22) {$\zeta = 9.1\times10^{-4}$}
\end{overpic}
\begin{overpic}[width=.32\textwidth, trim=0cm 0cm 0cm 0cm, clip]{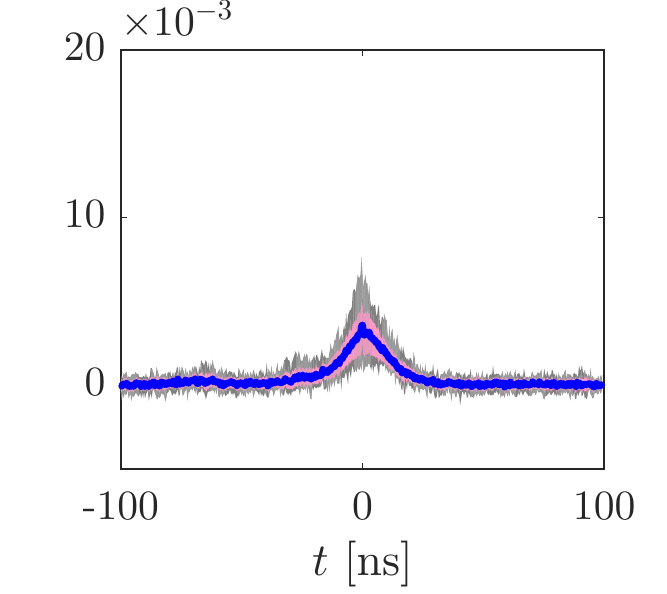}
\put (19,74) {(f)}
\put (58,74) {\footnotesize $\Delta p$=38kPa}
\put (57,66) {\footnotesize $R_{0}$=4.6mm}
\put (42,22) {$\zeta = 1.3\times10^{-3}$}
\end{overpic}
\begin{overpic}[width=.32\textwidth, trim=0cm 0cm 0cm 0cm, clip]{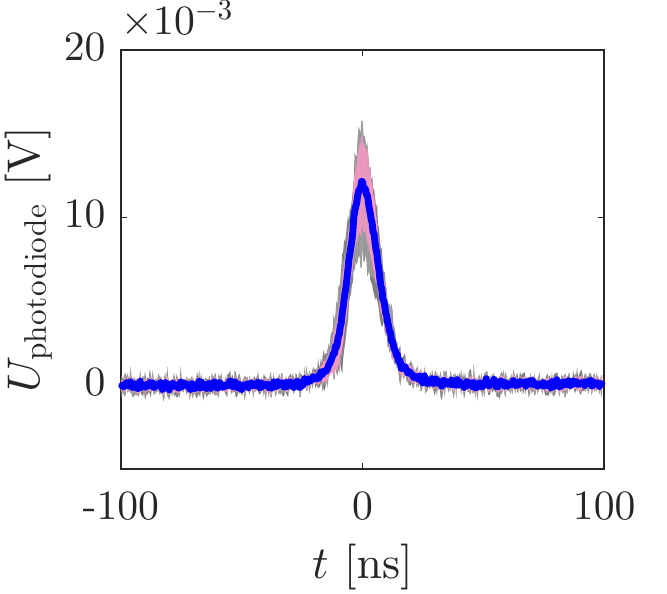}
\put (19,74) {(g)}
\put (58,74) {\footnotesize $\Delta p$=98kPa}
\put (57,66) {\footnotesize $R_{0}$=2.8mm}
\put (42,22) {$\zeta = 4.5\times10^{-4}$}
\end{overpic}
\begin{overpic}[width=.32\textwidth, trim=0cm 0cm 0cm 0cm, clip]{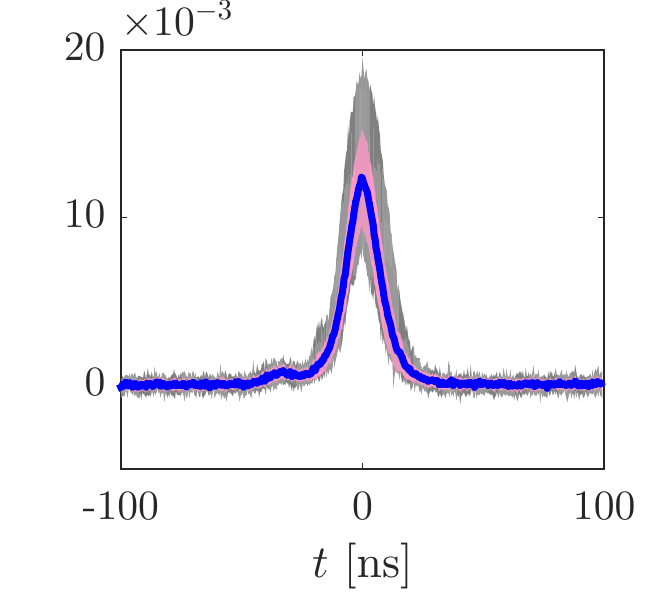}
\put (19,74) {(h)}
\put (58,74) {\footnotesize $\Delta p$=58kPa}
\put (57,66) {\footnotesize $R_{0}$=3.4mm}
\put (42,22) {$\zeta = 6.8\times10^{-4}$}
\end{overpic}
\begin{overpic}[width=.32\textwidth, trim=0cm 0cm 0cm 0cm, clip]{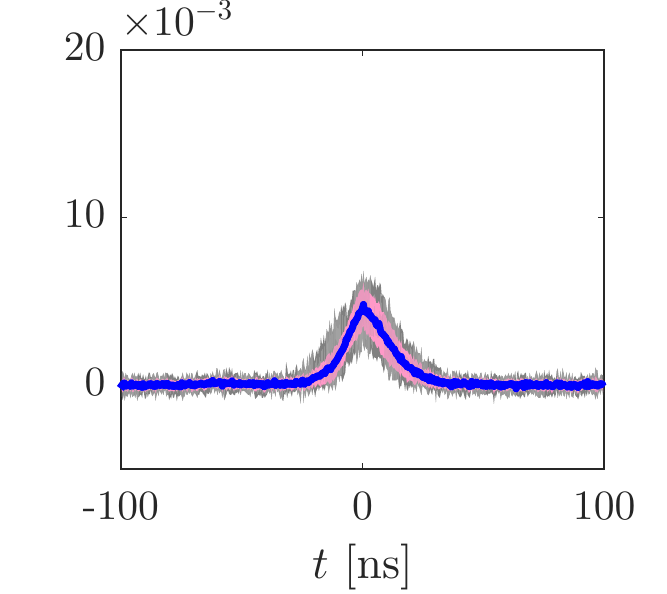}
\put (19,74) {(i)}
\put (58,74) {\footnotesize $\Delta p$=38kPa}
\put (57,66) {\footnotesize $R_{0}$=3.9mm}
\put (42,22) {$\zeta = 1.0\times10^{-3}$}
\end{overpic}
\begin{overpic}[width=.32\textwidth, trim=0cm 0cm 0cm 0cm, clip]{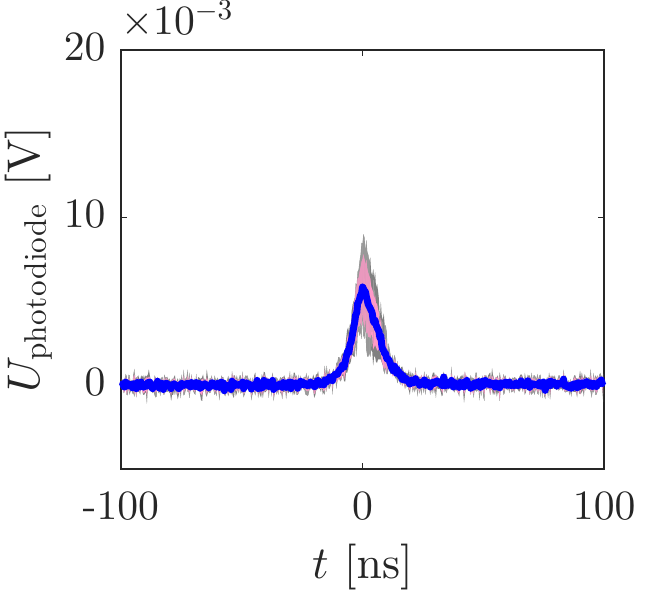}
\put (19,74) {(j)}
\put (58,74) {\footnotesize $\Delta p$=98kPa}
\put (57,66) {\footnotesize $R_{0}$=2.3mm}
\put (42,22) {$\zeta = 3.1\times10^{-4}$}
\end{overpic}
\begin{overpic}[width=.32\textwidth, trim=0cm 0cm 0cm 0cm, clip]{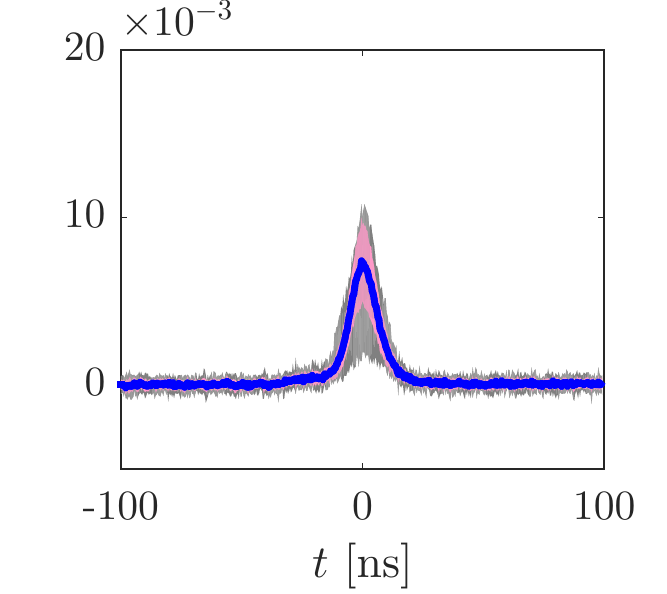}
\put (19,74) {(k)}
\put (58,74) {\footnotesize $\Delta p$=58kPa}
\put (57,66) {\footnotesize $R_{0}$=2.8mm}
\put (42,22) {$\zeta = 5.0\times10^{-4}$}
\end{overpic}
\begin{overpic}[width=.32\textwidth, trim=0cm 0cm 0cm 0cm, clip]{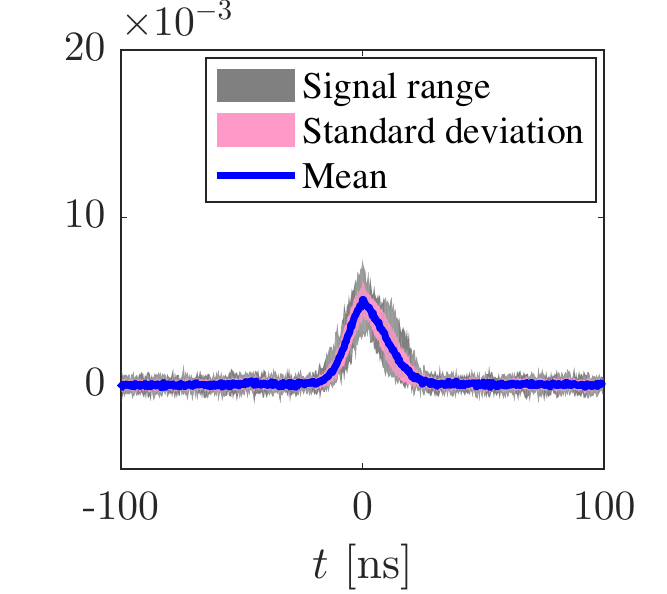}
\put (19,74) {(l)}
\put (58,52) {\footnotesize $\Delta p$=38kPa}
\put (57,44) {\footnotesize $R_{0}$=3.2mm}
\put (42,22) {$\zeta = 7.9\times10^{-4}$}
\end{overpic}
\caption{(Color online) Mean of 20 luminescence signals measured by the photodetector for different bubble energies, (a)-(c) $E_{0}=22$~mJ, (d)-(f) $E_{0}=15$~mJ, (g)-(i) $E_{0}=9$~mJ and (j)-(l) $E_{0}=5$~mJ, for three different driving pressures $\Delta p=98$~kPa, 58~kPa and 38~kPa. The range covered by the individual signals and the standard deviations are also displayed. $t=0$~ns corresponds to the maximum peak.}
\label{fig:shades}
\end{center}
\end{figure}

%
\begin{figure}
\begin{center}
\begin{overpic}[width=.48\textwidth, trim=0cm 0cm 0cm 0cm, clip]{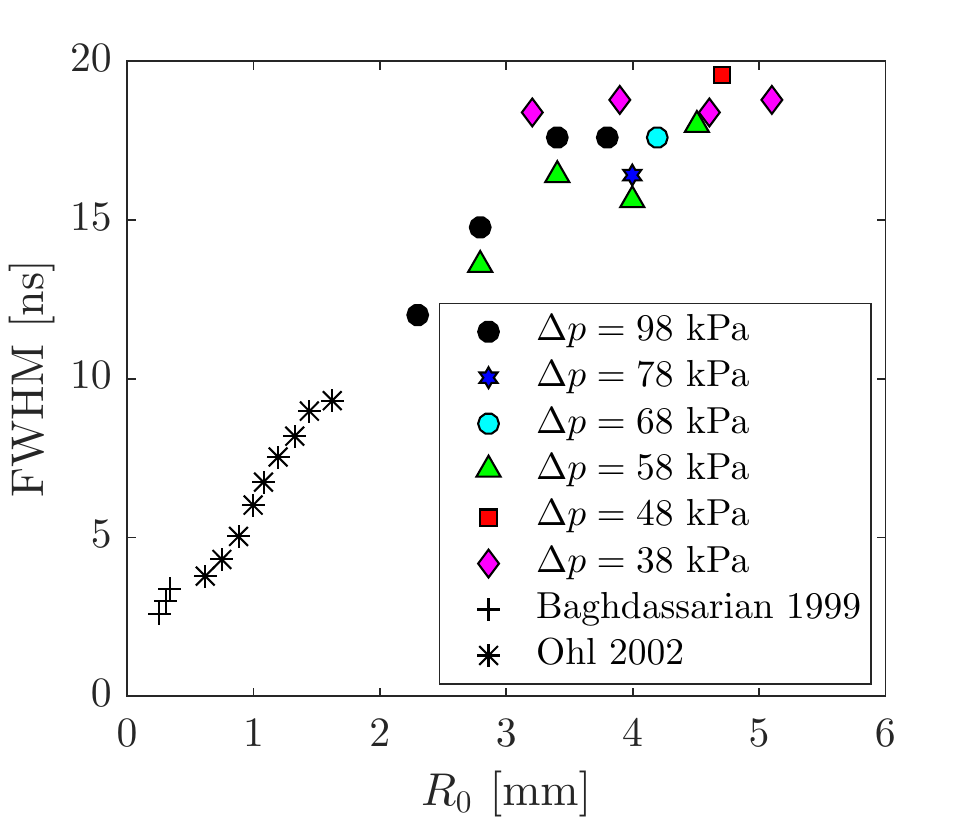}
\put (15,72) {(a)}
\end{overpic}
\begin{overpic}[width=.48\textwidth, trim=0cm 0cm 0cm 0cm, clip]{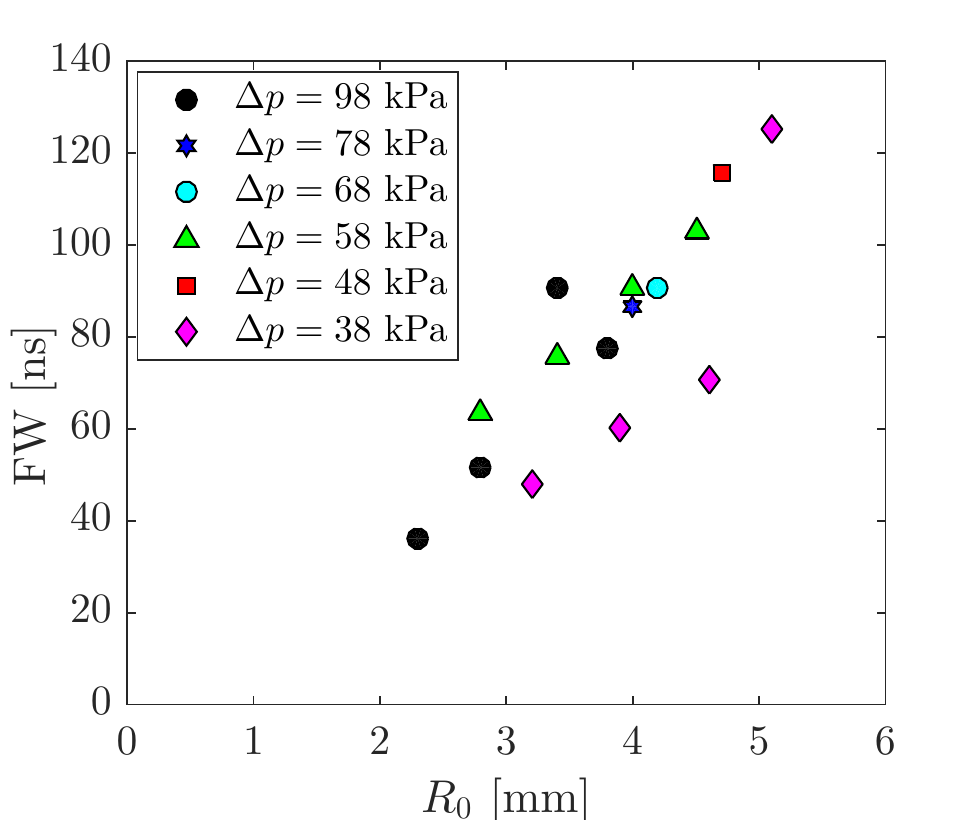}
\put (15,15) {(b)}
\end{overpic}
\caption{(Color online) (a) Full width at half maximum (FWHM) and (b) full width (FW, with 1\% of peak amplitude as threshold) of luminescence as a function of $R_{0}$. The durations have been extracted directly from the averaged photodetector signals of 20 bubbles. FWHM data at atmospheric pressure for reference are shown from Baghdassarian \emph{et al.} (1999)~\cite{Baghda1999} and from Ohl (2002)~\cite{Ohl2002}.}
\label{fig:R_dur}
\end{center}
\end{figure}

Figures~\ref{fig:shades}(a)--\ref{fig:shades}(l) show the averages of the photodetector signals at three different driving pressures ($\Delta p=98$, 58 and 38~kPa) and at four different bubble energies ($E_{0}=22$, 15, 9 and 5~mJ).
Each maximum peak is set to $t=0$~ns when the averaging is performed.
The range covered by the individual signals and the standard deviations are also displayed.
The more energetic bubbles show multiple peaks (Figs.~\ref{fig:shades}(a)--\ref{fig:shades}(f)), while at lower energies luminescence is measured as a single peak (Figs.~\ref{fig:shades}(g)--\ref{fig:shades}(l)).
Figures~\ref{fig:shades}(c) and \ref{fig:shades}(j) display signals with similar peak amplitudes, yet the high-energy bubble collapsing at low pressure yields multiple peaks while the low-energy bubble collapsing at atmospheric pressure yields a single peak.
Figures~\ref{fig:shades}(d) and \ref{fig:shades}(h) display signals for bubbles with the same maximum radius but with different energies, and, again, the higher-energy bubble yields more prominent additional peaks than the other.
However, we find no clear correlation between the number, amplitudes or timings of the peaks with the bubble's asphericity.

Figures~\ref{fig:R_dur}(a) and~\ref{fig:R_dur}(b) show the measured luminescence durations as the full width at half maximum (FWHM) and the full width (FW), which are extracted directly from the average of 20 individual photodetector signals.
The full width here is defined as the duration of the averaged signal above 1\% of its peak amplitude (the noise in the averaged signals have been smoothened out sufficiently not to affect this low threshold). 
In order to complete the graph for previously measured luminescence durations for smaller laser-induced bubbles, FWHM data from Baghdassarian \emph{et al.}~(1999)~\cite{Baghda1999} and from Ohl (2002)~\cite{Ohl2002} are included for purposes of comparison.
The trend for the duration of these large bubbles remains similar as for the previously reported smaller bubbles, that is, approximately linear as a function of $R_{0}$.
While past research has suggested the pulse duration to increase for bubbles collapsing at higher pressures~\cite{Brujan2005} and for bubbles deformed by a neighboring surface~\cite{Ohl2002}, the direct roles of $\Delta p$ and $\zeta$ on the pulse duration in our results are unclear.
In particular, luminescence durations at $\Delta p = 38$~kPa seem to be outliers from the general trend, with FWHM remaining almost constant for $R_{0}=3$--$5$~mm.

%
\begin{figure}
\begin{center}
\begin{overpic}[width=\textwidth, trim=0cm 0cm 0cm 0cm, clip]{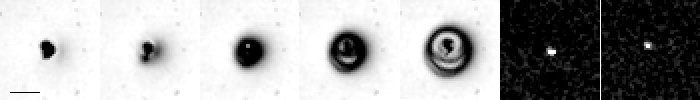}
\put (1,12) {(a)}
\put (72,12) {\textcolor{white}{(b)}}
\end{overpic}
\caption{Images of the luminescence emission of a bubble collapsing at $\Delta p=38$~kPa at normal gravity ($R_{0}=5.1$~mm, $\zeta=1.3\times10^{-3}$) captured by the ultra-high-speed CMOS camera, (a)~with a backlight LED and (b)~in the dark. The black bar shows the 1-mm scale. The interframe time is 100~ns and the exposure time is 50~ns. The contrast and brightness of the images have been adjusted to optimize the visual clarity of the events.}
\label{fig:imag_lum}
\end{center}
\end{figure}

Finally, a typical example of an ultra-high-speed CMOS camera recording of the luminescence is shown in Fig.~\ref{fig:imag_lum} where luminescence events are visible in the visualization with a backlight illumination (Fig.~\ref{fig:imag_lum}(a)) and in the dark (Fig.~\ref{fig:imag_lum}(b)) for a relatively deformed bubble.
Figure~\ref{fig:imag_lum}(b) shows the luminescent flash in the dark in two frames and thereby implies that the total luminescence events duration here exceeds the interframe time of 100~ns, consistent with the photodetector measurements (see Fig.~\ref{fig:R_dur}(b) for $R_{0}\approx 5$~mm).
The images here likely only capture the beginning and the end of the light emission, while the peak intensity occurs between the images (the exposure time 50~ns only covers half of the interframe time).
In fact, the CMOS camera systematically captures the luminescent flash in two or even three consecutive frames and occasionally gets saturated.
We also observe an upward shift of the light spot in the images of Fig.~\ref{fig:imag_lum}(b).
This might be expected, because according to momentum conservation, most of the bubble's translational motion upon its nonspherical collapse occurs during its last collapse and early rebound stages, when the luminescence is emitted.
The bubble centroid's upward displacement during the collapse is clearly visible in Fig.~\ref{fig:imag_lum}(a).

\section{Discussion}
\begin{figure}
\begin{center}
\begin{overpic}[width=.7\textwidth, trim=0cm 4.9cm 0cm 0.1cm, clip]{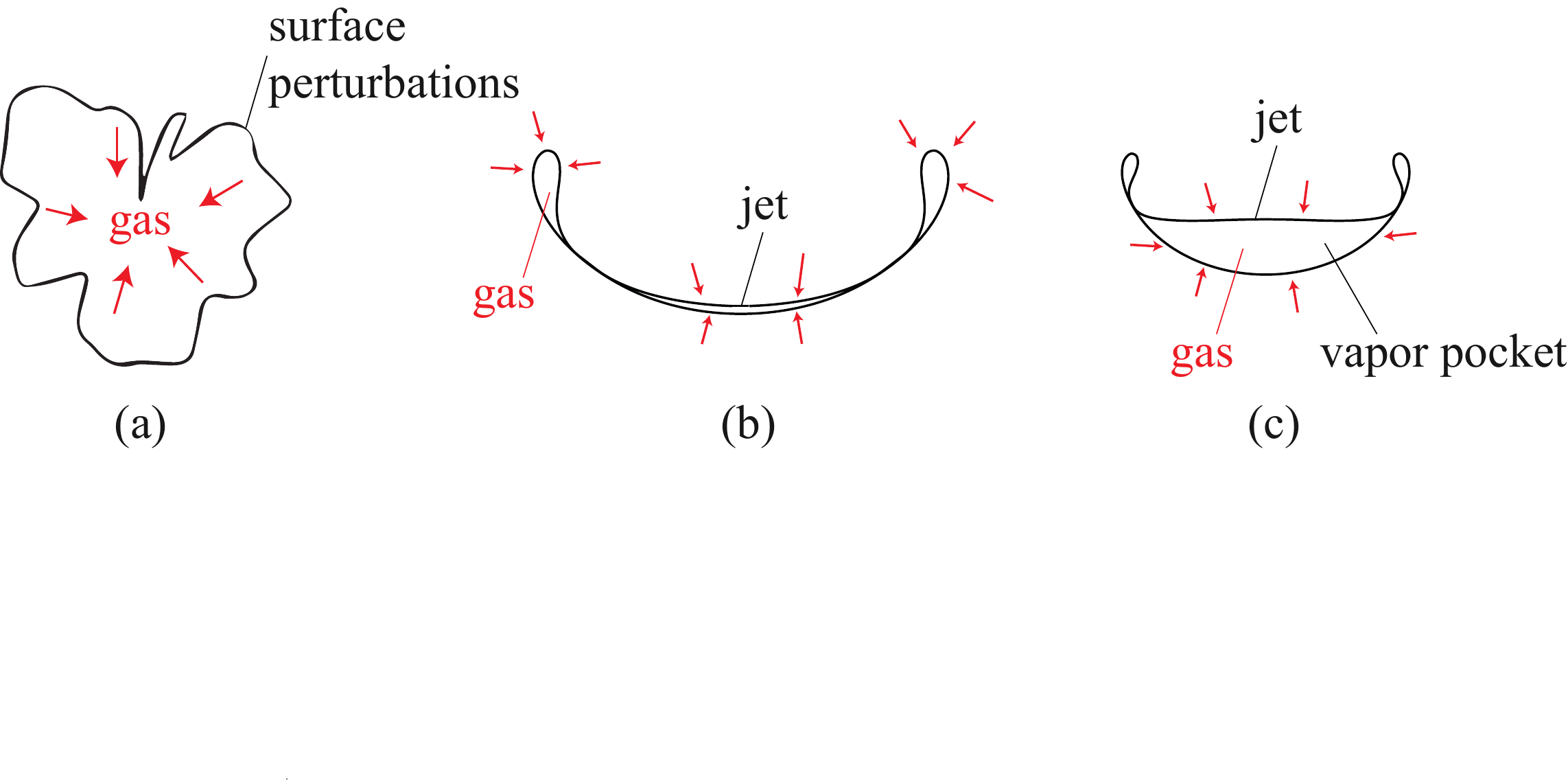}
\end{overpic}
\caption{(Color online) Illustration of the possible effect of the bubble's surface perturbations on its gas compression. Sketches of shapes at the final collapse stage for a bubble (a) with surface perturbations, (b) with a downward jet induced by a uniform pressure gradient and (c) with a downward jet induced by a neighboring free surface.}
\label{fig:ill_shapes}
\end{center}
\end{figure}

The results presented here give insight on how the topological changes of the cavity volume from a spherical to a jetting bubble affect the degree of adiabatic heating. 
Luminescence has an appreciable sensitivity on even the finest pressure field anisotropies in the liquid caused by the gravity-induced pressure gradient.
The threshold beyond which luminescence is no longer observed, $\zeta\approx3.5\times10^{-3}$, is close to the limit where we start observing jetting bubbles in our experiment ($\zeta\sim10^{-3}$), the latter however being a limit that is difficult to define with precision.
Considering a bubble deformation by nearby boundaries yielding an identical Kelvin impulse ($\gamma = \left(0.915\zeta^{-1}\right)^{1/2}$~\cite{Supponen2016}, where, equivalently, $\gamma=s/R_{0}$, $s$ being the distance between the bubble and the boundary), the threshold at which we no longer detect luminescence here would be equivalent to a bubble collapsing at a distance of 7.5 times its maximum radius from the boundary.
Such a limit is in disagreement with previous studies on luminescence from laser-induced bubbles deformed by near boundaries, where the equivalent limit is much lower, e.g.~$\gamma \sim 3.5$ in refs.~\cite{Ohl1998,Brujan2005} (corresponding to $\zeta \sim 0.016$). 
This discrepancy is possibly attributed either to different sources of deformations yielding different levels of gas compression, or to the sensitivity of luminescence on the initial bubble sphericity.
The latter hypothesis is supported by our previous observation that the level of deformation at which a microjet visibly pierces the bubble and drives a vapor-jet during the rebound for bubbles deformed by near surfaces in our experiment ($\zeta\approx10^{-3}$ or $\gamma \approx 14$) is also significantly lower compared to the literature (typically $\gamma \approx 5$)~\cite{Supponen2016}.
Likewise, we have recently measured the shock waves energy to start being sensitive to $\zeta$ at larger distances away from surfaces ($\gamma\approx8$~\cite{Supponen2017}) compared to the literature ($\gamma\approx3$~\cite{Vogel1988}).
As mentioned earlier, lens-based bubble generation systems, in contrast to the use of a high-convergence parabolic mirror, produce bubbles with higher surface perturbations that are amplified during the last collapse stage~\cite{Plesset1956}.
Consequently, a potential microjet, which can be regarded as the lowest-order deviation from a sphere and is thus most effective at inhibiting the final gas compression, may be masked by more important, higher-order perturbations.
This could make the bubble experience a collapse that perhaps more effectively compresses the gas and that is less susceptible to external factors, possibly even appearing spherical.
This hypothesis is illustrated in Fig.~\ref{fig:ill_shapes}.

%

\begin{figure}
\begin{center}
\begin{overpic}[width=.5\textwidth, trim=0cm 0cm 0cm 0cm, clip]{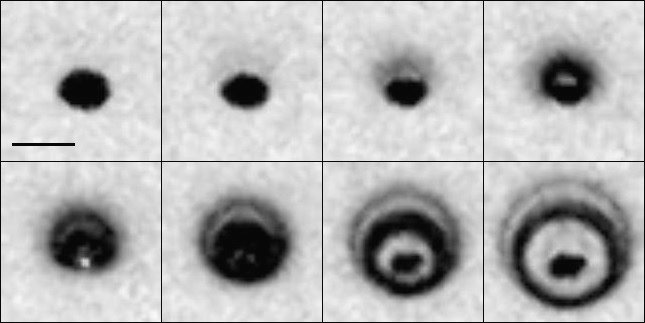}
\end{overpic}
\caption{Luminescence emission at the location of jet impact for a bubble collapsing near a free surface. $R_{0}=4.1$~mm, $\gamma=7.2$. The interframe time is 100~ns and the black bar shows the 1-mm scale. The microjet is directed downwards. Adapted from Ref.~\cite{Supponen2017}.}
\label{fig:lumtip}
\end{center}
\end{figure}

However, luminescence can also occur for jetting bubbles, as has previously been shown for bubbles deformed by a neighboring surface~\cite{Ohl1998}, for acoustic cavitation bubbles in multibubble fields in xenon saturated phosphoric acid~\cite{Cairos2017} and for xenon gas bubbles collapsed by a passing shock wave~\cite{Kappus2011}.
A possible reason for us not to observe light emission for bubbles that produced clear gravity-driven `vapor-jets' upon rebound could be linked to the characteristic shape that the bubble assumes at the moment of the jet piercing.
We have previously shown that, according to potential flow theory, the gravity-induced deformation yields a broad jet whose shape is highly similar to the one of the bubble wall it pierces~\cite{Supponen2016}, and thereby the gas compression after the jet impact becomes particularly weak (see Fig.~\ref{fig:ill_shapes}(b)).
In contrast, when the bubble is deformed by a neighboring rigid or a free surface, at certain ranges of $\zeta$, potential flow theory predicts small vapor `pockets' remaining between the jet and the opposite bubble wall upon first contact of jet onto it~\cite{Supponen2016}, such as in the illustration of Fig.~\ref{fig:ill_shapes}(c).
We have previously observed luminescence from the location at which the jet pierces the bubble wall for bubbles collapsing near a free surface, as shown in Fig.~\ref{fig:lumtip} (adapted from Ref.~\cite{Supponen2017}).
This is due to the contact between the jet and the opposing wall being more irregular, which is characteristic to bubbles near free surfaces. 
The jet thus divides the bubble into multiple separate segments, one of which is a vapor pocket between the jet and the opposite wall that is individually able to collapse in an almost spherical way, which, in turn, yields an effective compression.
This hypothesis is supported by our previous observations where such vapor pockets emitted strong shocks for bubbles near a free surface~\cite{Supponen2017}.
However, we are unable to temporally distinguish the jet impact from the individual collapses of the remaining bubble segments at a low enough $\zeta$ for luminescence to still be visible.
It would be interesting in the future to study more thoroughly the effect of the bubble shape on luminescence by varying this shape by different sources of deformation (e.g.~comparing different surfaces and gravity) in a single setup.

%

A surprising finding is that the spectroscopically estimated blackbody temperatures of luminescence barely vary with the different levels of bubble deformation (Fig.~\ref{fig:zE_p}(b)), while its energy varies by two orders of magnitude (Fig.~\ref{fig:zE_p}(a)).
We do not exclude the possibility that the scatter of the data, partly caused by the fitting error, hides a possible weak variation of the blackbody temperature with $\zeta$.
However, it could also be due to the fact that as the radiation power scales as $T_{\rm lum}^{4}$, any attempt to increase $T_{\rm lum}$ immediately results in an accelerated loss of energy by radiation.
Another potential physical reason could be the presence of water vapor which increases the heat capacity ratio~\cite{Storey2000,Toegel2000,Brenner2002}.
It has been shown numerically that sonoluminescent bubbles that have compression ratios beyond $R_{0}/R_{\rm min}\sim 20$, water vapor starts affecting the power-law increase of the maximum temperature with the compression ratio, finally asymptoting to $T_{\rm lum}\approx10000$~K~\cite{Storey2000}.
It is difficult to measure the minimum bubble size in our experiment due to the luminescence and the light deflection caused by the pressure rise in the surrounding liquid `hiding' the bubble at the last stage of the collapse (see images in Fig.~\ref{fig:lum}).
However, when choosing the luminescent flash size as the minimum radius, we get compression ratios of $R_{0}/R_{\rm min}> 40$, which is already in the regime where vapor affects the heating.

\begin{figure}
\begin{center}
\begin{overpic}[width=.5\textwidth, trim=0cm 0cm 0cm 0cm, clip]{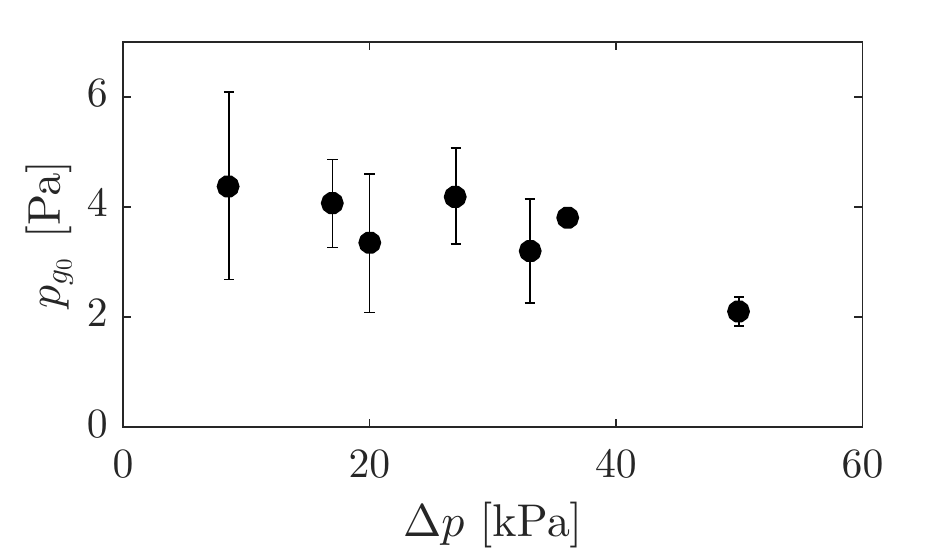}
\end{overpic}
\caption{Averaged initial partial pressure of the noncondensible gas, estimated by fitting the Keller-Miksis model to the observed rebound radial evolution, as a function of driving pressure $\Delta p$. The data contain bubbles of different radii (1-3.5~mm) collapsing highly spherically ($\zeta<0.0007$), and the error bars show the standard deviation.}
\label{fig:pg0}
\end{center}
\end{figure}

The noncondensible gas trapped inside the bubble plays a key role on luminescence emission.
We believe the bubble contains 1)~vapor, of which the partial pressure is assumed to stay at the liquid vapor pressure $p_{v}$ during most of the bubble's lifetime; 2)~the laser-generated gas (demonstrated in Ref.~\cite{Sato2013}), which we assume to depend on the energy deposited by the laser to generate the bubble, that is, proportional to $E_{\rm laser}\propto E_{0}\propto R_{0}^{3}\Delta p$; and 3)~the diffused gas from the water to the bubble, which depends both on the total bubble surface during its lifetime, which is proportional to $R_{0}^{3}\Delta p^{1/2}$, and on the diffusion-driving pressure $\Delta p$.
Each of these likely contribute to the noncondensible gas, which is difficult to measure directly.
The laser-generated and diffused gas are both proportional to the bubble's maximum volume, while they may depend on $\Delta p$ to a different extent.
A method has been proposed by Tinguely~\emph{et al.}~\cite{Tinguely2012} to estimate the initial partial pressure of the noncondensible gas $p_{g_{0}}$ by fitting the Keller-Miksis model~\cite{Keller1980} to the observed rebound.
Applying this method on the observed radial evolution of spherically collapsing bubbles at various $\Delta p$ in microgravity, we can estimate the variation of $p_{g_{0}}$ as a function of $\Delta p$.
Our preliminary results find that $p_{g_{0}}$ remains almost constant ($p_{g_{0}}\approx4$~Pa) for the range of $\Delta p$ covered here, deferring less than the standard deviation, as illustrated in Fig.~\ref{fig:pg0}.
Furthermore, the luminescence energy data obtained in this range of $\Delta p$ in Fig.~\ref{fig:zE_p} suggest that the bubble's deformation ($\zeta$) is the dominant source of luminescence energy hindering rather than $\Delta p$, even if a weak dependence on the latter may exist.
Figure~\ref{fig:R_Elum}, that shows luminescence energies as a function of the bubble energy at different $\Delta p$, however, suggests some additional dependence of luminescence energy on $\Delta p$.
A systematic study with a controlled gas content of the water, preferably in microgravity to remove the effect of bubble's deformation by gravity, would be useful to clarify the effect of noncondensible gas on luminescence and on other bubble collapse phenomena.

%

Finally, it would be interesting to understand the physics behind the multiple luminescence emission events that are measured by the photodetector (Figs.~\ref{fig:water1} and~\ref{fig:shades}).
These peaks show considerable fluctuations in their number, amplitudes, shapes and timings.
The timing of the strongest luminescence event with respect to the emission of the collapse shock wave is remarkably reproducible, varying by only $\sim10$~ns (Fig.~\ref{fig:water1}).
The finding that larger bubbles emit more peaks than smaller ones is consistent with the literature, although the bubble sizes reported in the past were much smaller overall and multiple peaks were observed for bubbles with $R_{0} < 2$~mm~\cite{Ohl2002}.
The discrepancy between our observations (single peak for $R_{0}<3$~mm) and the past literature is, again, likely due to the high initial sphericity of the bubble in our experiment.
The multiple peaks could be associated with different hot spots that could be a result of an inhomogeneous bubble interior or bubble splitting, as suggested by Ohl~\cite{Ohl2002}, which would indeed be strongly affected by the initial bubble sphericity.
They could also be linked to plasma instabilities, to minor impurities trapped within the bubble or the potential formation of a `hidden' (non-piercing) microjet, which is challenging to verify since the levels of deformations here are so weak.

\section{Conclusion}

In this work, we have captured broad spectra of single cavitation bubble luminescence from individual collapses using an innovative measurement technique.
We have measured luminescence from a previously uncovered range of maximum radii ($R_{0}=1.5$--$6$~mm) of laser-induced bubbles, thanks to their high initial sphericity.
The bubbles were controllably deformed from highly spherical to jetting bubbles under the effect of the gravity-induced hydrostatic pressure gradient.
The deformation was quantified with the dimensionless anisotropy parameter $\zeta$, which was adjusted via maximum bubble radius, driving pressure and variable gravity aboard parabolic flights.
We found a rapid decrease of the relative luminescence energy $E_{\rm lum}/E_{0}$ with $\zeta$.
No clear variation of the fitted blackbody temperature, which ranged between $T_{\rm lum}=7000$ and 11500~K, as a function of $\zeta$ or the driving pressure was found.
The threshold of luminescence approximately coincides with $\zeta$ at which we start observing vapor-jets in our experiment.
The light emission is found to be nonuniform in time for the most energetic bubbles, as multiple events are detected in the time-resolved measurements by a photodetector, while low-energy bubbles emitted single luminescence peaks.
The luminescence events were found to occur in a time frame of 200~ns. 
The full width at half maximum of the averaged luminescence signal scales with $R_{0}$ and is generally on the order of 10--20~ns.

\acknowledgements{We gratefully acknowledge the support of the Swiss National Science Foundation (Grant no. 200020--144137), the University of Western Australia Research Collaboration Award, and European Space Agency for the 60th and 62nd parabolic flight campaigns and Professor Oliver Ullrich for the 1st Swiss parabolic flight. We thank Nicolas Dorsaz, Marc Tinguely and Christophe Praz for their valuable help with the experiment.}

\bibliography{luminescence}

\end{document}